\documentclass[twocolumn,prb,aps,superscriptaddress]{revtex4-2}
\usepackage{amsmath}
\usepackage{amssymb}
\usepackage{bm}
\usepackage{graphicx}
\usepackage{dcolumn}
\usepackage{graphicx}
\usepackage{epstopdf}
\usepackage{epsfig}
\usepackage{color}

\begin{document}
\title{Crystal Electic Field and Possible Coupling with Phonon in Kondo lattice CeCuGa${\bm _3}$}

\author{V.\ K.\ Anand}
\altaffiliation{vivekkranand@gmail.com}
\affiliation{ISIS Neutron and Muon Facility, STFC, Rutherford Appleton Laboratory, Chilton, Oxfordshire, OX11 0QX, UK}
\affiliation{Department of Physics, University of Petroleum and Energy Studies, Dehradun, Uttarakhand, 248007, India}
\author{A. Fraile}
\affiliation{Nuclear Futures Institute, Bangor University, Bangor, LL57 1UT, UK}
\author{D. T. Adroja}
\altaffiliation{devashibhai.adroja@stfc.ac.uk}
\affiliation{ISIS Neutron and Muon Facility, STFC, Rutherford Appleton Laboratory, Chilton, Oxfordshire, OX11 0QX, UK}
\affiliation{\mbox{Highly Correlated Matter Research Group, Physics Department, University of Johannesburg, P.O. Box 524,} Auckland Park 2006, South Africa}
\author{Shivani Sharma}
\affiliation{ISIS Neutron and Muon Facility, STFC, Rutherford Appleton Laboratory, Chilton, Oxfordshire, OX11 0QX, UK}
\author{Rajesh Tripathi}
\affiliation{ISIS Neutron and Muon Facility, STFC, Rutherford Appleton Laboratory, Chilton, Oxfordshire, OX11 0QX, UK}
\affiliation{Jawaharlal Nehru Centre for Advanced Scientific Research, Jakkur, Bangalore 560064, India}
\author{\mbox{C. Ritter}}
\affiliation{Institut Laue-Langevin, Boite Postale 156, 38042 Grenoble Cedex, France}
\author{C. de la Fuente}
\affiliation{\mbox{Departamento de Fisica de Materia Condensada, Facultad de Ciencias, Universidad de Zaragoza,} E-50009 Zaragoza, Spain}
\author{P. K. Biswas}
\author{V. Garcia Sakai}
\affiliation{ISIS Neutron and Muon Facility, STFC, Rutherford Appleton Laboratory, Chilton, Oxfordshire, OX11 0QX, UK}
\author{A. del Moral}
\affiliation{\mbox{Departamento de Fisica de Materia Condensada, Facultad de Ciencias, Universidad de Zaragoza,} E-50009 Zaragoza, Spain}
\author{A. M. Strydom}
\affiliation{\mbox{Highly Correlated Matter Research Group, Physics Department, University of Johannesburg, P.O. Box 524,} Auckland Park 2006, South Africa}

\date{\today}

\begin{abstract}
We investigate the magnetic and crystal electric field (CEF) states of the Kondo lattice system CeCuGa$_{3}$ by muon spin relaxation  ($ \mu$SR), neutron diffraction, and inelastic neutron scattering (INS) measurements. A noncentrosymmetric BaNiSn$_3$-type tetragonal crystal structure (space group $I4\,mm$) is inferred from x-ray as well as from neutron powder diffraction. The low-temperature magnetic susceptibility and heat capacity data show an anomaly near 2.3\,-\,2.5~K associated with long range magnetic ordering, which is further confirmed by $ \mu$SR and neutron diffraction data. The neutron powder diffraction collected at 1.7~K shows the presence of magnetic Bragg peaks indexed by an incommensurate magnetic propagation vector {\bf k}~$ = (0.148, 0.148, 0)$ and the magnetic structure is best described by a longitudinal spin density wave with ordered moments lying in $ab$-plane. An analysis of the INS data based on a CEF model reveals the presence of two magnetic excitations near 4.5~meV and 6.9~meV. The magnetic heat capacity data suggest an overall CEF splitting of 20.7~meV, however the excitation between 20 and 30~meV is very broad and weak in our INS data, but could provide an evidence of CEF level in this energy range in agreement with the magnetic entropy. Our analysis of INS data based on the CEF-phonon model indicates that the two excitations at 4.5 meV and 6.9 meV have their origin in CEF-phonon coupling (i.e. splitting of one CEF peak into two peaks, called vibron), with an overall splitting of 28.16~meV,  similar to the case of CeCuAl$_3$ and CeAuAl$_3$. 
\end{abstract}

\maketitle

\section{\label{Intro} INTRODUCTION}

Ce-based intermetallic compounds have been very appealing for their intriguing properties owing to the strongly competing Ruderman-Kittel-Kasuya-Yosida (RKKY) and Kondo interactions.\cite{Stewart1984,Riseborough, Amato, Lohneysen, Pfleiderer2009, Si2010} They present a wide variety of electronic ground states including magnetic ordering, superconductivity, valence fluctuations, heavy-fermion and quantum critical behavior. For example, the Kondo lattice heavy fermion systems CeRhSi$_{3}$ and CeIrSi$_{3}$ exhibit long-range antiferromagnetic ordering and undergo pressure-induced superconductivity.\cite{Kimura2005,Kimura2007, Sugitani2006, Okuda2007} Interestingly, they have noncentrosymmetric crystal structures (BaNiSn$_3$-type tetragonal, space group $I4\,mm$) due to which the superconductivity in these Ce$TX_3$ compounds is unconventional in nature as it allows mixing between the spin singlet and spin-triplet parities and hence mixed parity.\cite{Bauer2012}

Recently our group has been working on $RTX_3$ ($R$ = rare earth, $T$ = transition metal and $X$ = Si, Ge, Al, Sn, Ga) compounds \cite{Anand2011a,Anand2011b, Anand2011c, Anand2012a, Adroja2012a, Hillier2012, Anand2013, Smidman2013, Anand2014a, Anand2014b, Smidman2014, Adroja2015, Anand2016, Anand2018, Adroja2012,Fuente2021} to search for novel materials and understand the magnetism, superconductivity and crystal electric field (CEF) effects in $RTX_3$, in particular, using neutron and muon techniques. In our efforts, very recently some of us investigated the crystal electric field (CEF) excitations in noncentrosymmetric tetragonal CeCuAl$_{3}$  using  inelastic neutron scattering (INS) measurements and found the presence of three magnetic excitations.\cite{Adroja2012} For Ce$^{3+}$ ($J = 5/2$), being a Kramers ion, according to Kramer’s degeneracy theorem, in the paramagnetic state one would expect only two CEF excitations from the ground state. Therefore the observation of three CEF excitations in CeCuAl$_{3}$ had been very exciting, providing evidence of strong CEF-phonon coupling (under a magneto-elastic mechanism) in this compound.\cite{Adroja2012} Similarly, the inelastic neutron scattering study on CeAuAl$_{3}$ single crystal also revealed the presence of CEF-phonon coupling.\cite{Petr2019} Motivated by these observations, we have performed an INS investigation of CEF excitations in the Kondo lattice CeCuGa$_{3}$.

The physical properties of CeCuGa$_{3}$ have been investigated by different groups, and depending on the stoichiometry of Cu and Ga, a sample dependent ground state has been found.\cite{Sampathkumaran1992, Mentink1993, Martin1996, Aoyama1996, Martin1998, Oe2010, Kontani1999, Joshi2012,Przybylski2014}  Mentink {\it et al}. \cite{Mentink1993} reported a paramagnetic behavior down to 0.4 K and found evidence for Kondo lattice heavy fermion behavior in polycrystalline CeCuGa$_{3}$ having BaNiSn$_{3}$-type noncentrosymmetric tetragonal structure (space group $I4\,mm$). On the other hand, Martin {\it et al}. suggested an antiferromagnetic ordering below 1.9~K in BaNiSn$_{3}$-type polycrystalline CeCuGa$_{3}$.\cite{Martin1996}  Neutron diffraction (ND) study on polycrystalline noncentrosymmetric CeCuGa$_{3}$ revealed an incommensurate magnetic structure with a magnetic propagation vector ${\bf k}=(0.176, 0.176, 0)$ and a helical arrangement of moments.\cite{Martin1998} For single crystal CeCuGa$_{3}$, in addition to the magnetic Bragg peaks [indexed by {\bf k} = (0.176, 0.176, 0)], the ND data also revealed the presence of satellite peaks indexed by an incommensurate propagation vector of ${\bf k}_{\rm s}=(0.137, 0.137, 0)$ which persists up to 300 K and is suggested to be the result of a short-ranged structural modulation of the crystal structure \cite{Martin1998}. Oe {\it et al}. found evidence for ferromagnetic ordering in single crystal CeCu$_{0.8}$Ga$_{3.2}$ having BaAl$_4$-type ($I4/mmm$) crystal structure. \cite{Oe2010} Joshi {\it et al}. also found a ferromagnetic ground state in single crystal CeCuGa$_{3}$ which formed in disordered ThCr$_{2}$Si$_{2}$-type tetragonal structure (space group $I4/mmm$).\cite{Joshi2012} It has been conjectured that the ground state properties are controlled by the degree of Cu-Ga disorder which in turn leads to two variants of crystal structure: ThCr$_{2}$Si$_{2}$-type tetragonal structure ($I4/mmm$) and BaNiSn$_{3}$-type tetragonal structure ($I4\,mm$, a  noncentrosymmetric structure). According to Joshi {\it et al}. CeCuGa$_{3}$ forming in disordered ThCr$_{2}$Si$_{2}$-type tetragonal structure ($I4/mmm$) exhibits a ferromagnetic ground state.\cite{Joshi2012} The structure dependent magnetic properties can be correlated to the different atomic environments for disordered centrosymmetric ($I4/mmm$) and ordered noncentrosymmetric ($I4\,mm$) structures.

We have investigated the CEF excitations in polycrystalline CeCuGa$_{3}$ sample forming in noncentrosymmetric BaNiSn$_{3}$ structure ($I4\,mm$) which we report here. CEF excitations in CeCuGa$_{3}$ have not been previously investigated by INS measurements. Oe {\it et al}.\ and Joshi {\it et al}.\ extracted information about CEF parameters by fitting the single crystal susceptibility data in the paramagnetic temperature range of ferromagnetically ordered CeCuGa$_{3}$ ($I4/mmm$).\cite{Oe2010,Joshi2012} From INS data we find two strong CEF excitations at 4.5~meV and 6.9~meV. Our CEF results substantially differ from those obtained from the fitting of magnetic susceptibility by Joshi {\it et al}.\ who found an overall CEF splitting of 228~K (19.6~meV) with a first excited CEF doublet at 50~K (4.3~meV).\cite{Joshi2012} Our magnetic heat capacity also reveals an overall CEF splitting of 20.7~meV. Apparently, the two excitations seen in INS (below 7~K) do not seem to originate from single-ion CEF transitions, but rather result from CEF-phonon coupling as found in CeCuAl$_{3}$ \cite{Adroja2012} and CeAuAl$_{3}$.\cite{Petr2019} Furthermore, our $ \mu$SR study finds a long range magnetic ordering in CeCuGa$_{3}$, and the neutron powder diffraction data reflect a longitudinal spin density wave with ordered moments in the (1 1 0) direction.

\section{\label{Exp} Experimental}

The polycrystalline samples of CeCuGa$_{3}$ as well as the nonmagnetic analog LaCuGa$_{3}$ ($\sim 26$~g each) were prepared by the standard arc-melting of high purity elements (Ce and La 99.9\% and Cu and Ga 99.99\%) in stoichiometric ratios. In order to improve the homogeneity and reaction among the constituent elements, the samples were flipped and re-melted several times during the arc-melting process. The resulting ingots were subsequently annealed at 900$^\circ$~C for 7 days in vacuum of $\sim 10^{-6}$ Torr. The BaNiSn$_{3}$-type crystal structure and the phase purity of the annealed sample were checked by the x-ray powder diffraction (XRPD) using the copper K$_{\alpha}$ radiation. Magnetization $M$ versus temperature $T$ and $M$ versus magnetic field $H$ measurements were made using a commercial superconducting quantum interference device (SQUID) magnetometer (MPMS, Quantum Design Inc.). The heat capacity $C_{\rm p}(T)$ measurements were made using the heat capacity option of a physical properties measurements system (PPMS, Quantum Design Inc.).

The $\mu$SR measurements were carried out in zero field using the MuSR spectrometer at the ISIS facility, Didcot, U.K\@. The powdered CeCuGa$_{3}$ sample was mounted on a high purity silver plate using diluted GE varnish and cooled inside the standard He-4 cryostat. The $\mu$SR data were collected at several temperatures between 1.2~K to 10~K. The neutron diffraction (ND) measurements were carried out on powdered CeCuGa$_{3}$ sample using the D20 powder neutron diffractometer at the Institute Laue Langevin, Grenoble, France. The powder sample was filled in a 10~mm vanadium can and cooled down to 1.7~K using a standard He-4 cryostat. The ND data were collected at 10~K and 1.7~K with the neutron beam of wavelength $\lambda = 2.41$~{\AA}. The ND data were analyzed by using the package FullProf.\cite{Rodriguez1993}

The INS experiments on CeCuGa$_{3}$ and LaCuGa$_{3}$ were performed on the MARI time of flight (TOF) spectrometer at the UK ISIS Neutron Spallation source. The powdered samples of these materials were wrapped in thin Al-foils and mounted inside thin-walled cylindrical Al-cans. Low temperatures down to 4.5 K were obtained by cooling the sample mounts in a top-loading closed cycle refrigerator with He-exchange gas. The INS data were collected for scattering angles between 3$^\circ$ and 135$^\circ$ using neutrons with incident energies $E_i = 8$, 15, and 40~meV. The data are presented in absolute units, mb/meV/sr/f.u.\ by using the absolute normalization obtained from the standard vanadium sample measured in identical conditions.  Low energy INS data were collected at 6~K using the  time-of-flight inverted-geometry crystal analyzer spectrometer OSIRIS with a PG002 analyzer and selecting the final neutron energy of 1.845~meV at the ISIS Facility. The elastic resolution (FWHM) was $17.5~\mu$eV.

\section{\label{MT-HC} Magnetic Susceptibility and Heat Capacity}

\begin{figure} 
\includegraphics[width=\columnwidth]{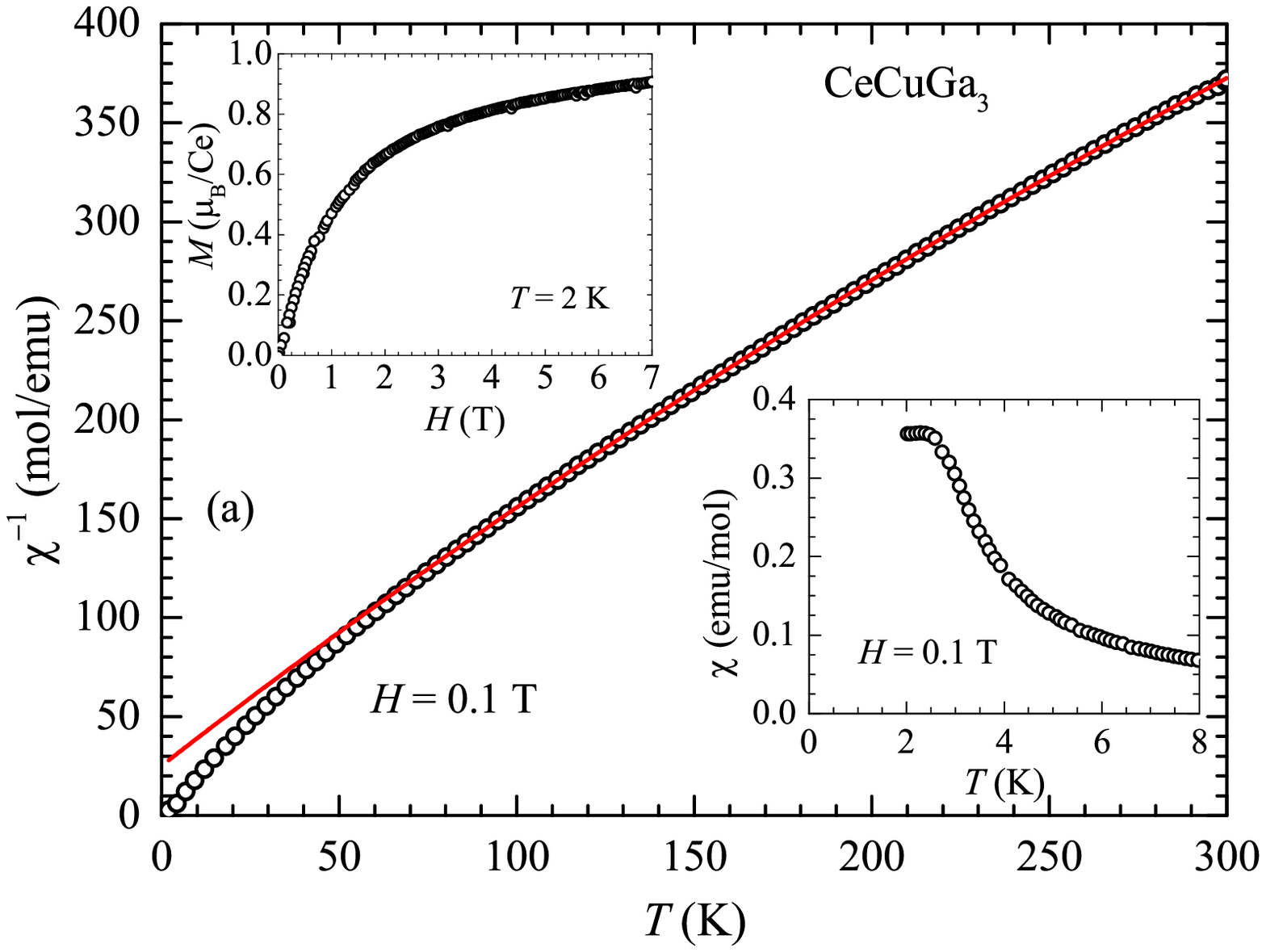}\vspace{0.05cm}
\includegraphics[width=\columnwidth]{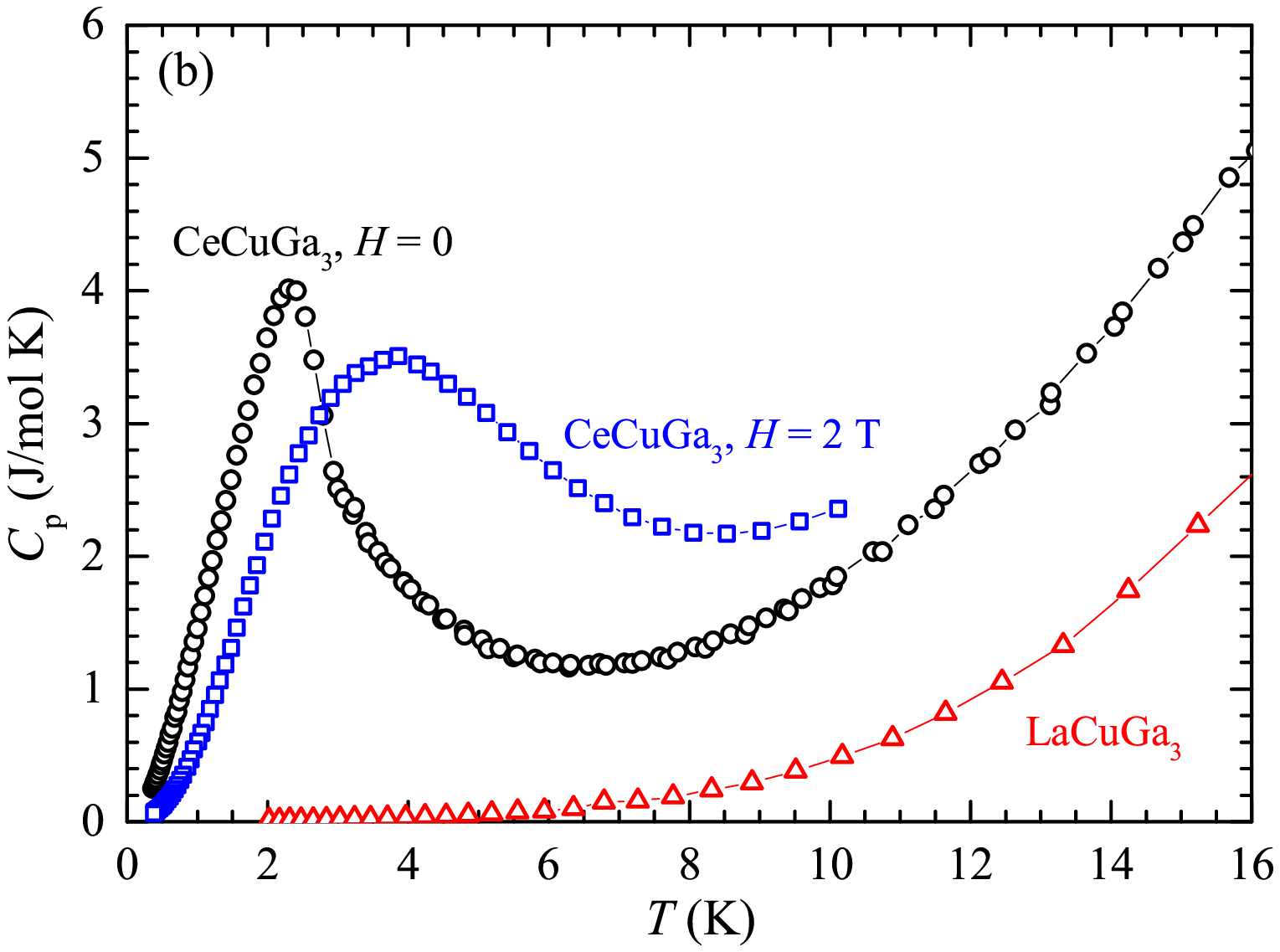}\vspace{0.05cm}
\includegraphics[width=\columnwidth]{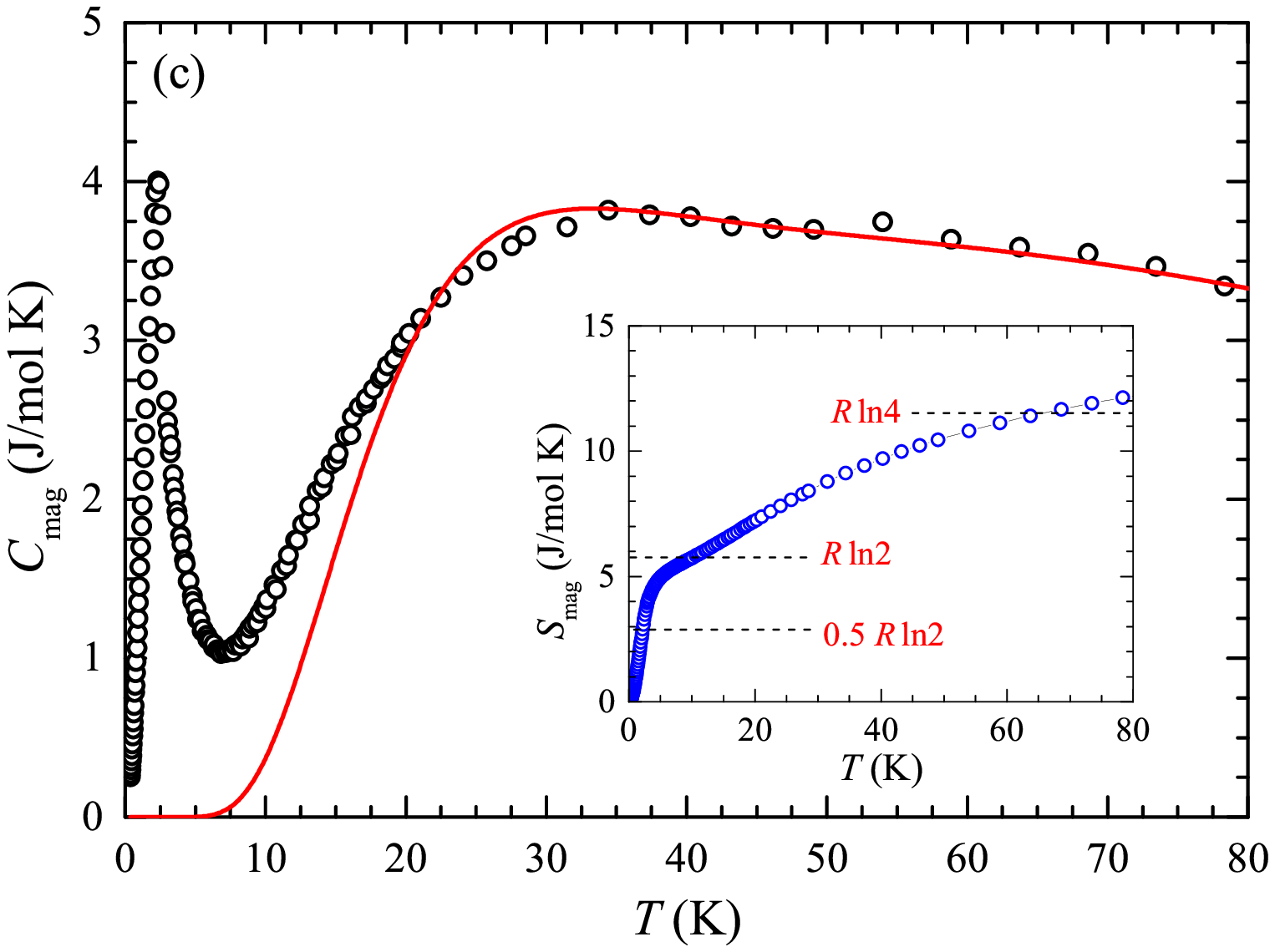}
\caption{\label{fig:MT-MH} (a) Magnetic susceptibility $\chi$ of CeCuGa$_{3}$ measured in a field of $H = 0.1$~T as a function of temperature $T$ plotted as $\chi^{-1}(T)$. The solid line represents the fit of $\chi^{-1}(T)$ data by modified Curie-Weiss law over 100~K to 300~K\@. Lower inset shows the low-$T$ $\chi(T)$ data. Upper inset shows the isothermal magnetization $M(H)$ measured at 2 K\@. (b) Heat capacity $C_{\rm p}(T)$ of CeCuGa$_{3}$ and non-magnetic reference LaCuGa$_{3}$ measured in $H = 0$ and $H = 2$~T\@. (c) Magnetic contribution to heat capacity $C_{\rm mag}(T)$. Solid red curve represents the crystal electric field contribution. Inset shows the magnetic entropy $S_{\rm mag}(T)$.}
\end{figure}

Figure~\ref{fig:MT-MH}(a) shows the temperature $T$ dependence of the inverse magnetic susceptibility $\chi^{-1}$ for CeCuGa$_{3}$ measured in a field of $H = 0.1$~T\@. The low-$T$ $\chi(T)$ data reveal an anomaly near 2.5~K (lower inset of Fig.~\ref{fig:MT-MH}(a)) indicating a magnetic phase transition. The isothermal magnetization $M(H)$ measured at 2 K presented in the upper inset of Fig.~\ref{fig:MT-MH}(a) shows a nonlinear behavior and a tendency of near saturation however with reduced $M$ of $0.9\,\mu_{\rm B}$ at 2~K and 7~T\@. The near saturation magnetization is much lower than the theoretically expected saturation value of $M_s=g_J J \mu_{\rm B} = 2.14\, \mu_{\rm B}$ for Ce$^{3+}$ ions. The reduction in $M$ can be attributed to combined effects of both Kondo and crystal electric field.

At high-$T$ the $\chi(T)$  data follow a modified Curie-Weiss behavior $\chi = \chi_0 + C/(T-\theta_{\rm p})$. A fit of the $\chi^{-1}(T)$ data over 100~K to 300~K (solid red curve in Fig.~\ref{fig:MT-MH}(a)) yields $\chi_0 = 4.8\times 10^{-4}$ emu/mol, an effective moment of $2.37 \,\mu_{\rm B}$ and a Weiss temperature $\theta_{p} = -18(2)$~K\@. The effective moment so obtained is close to the theoretically expected value of $2.54 \,\mu_{\rm B}$ for Ce$^{3+}$ ions. The negative value of $\theta_{p}$ is indicative of dominant antiferromagnetic interaction. From the value of $\theta_{p}$ we estimate the Kondo temperature using the relation $T_{\rm K} = |\theta_{p}|/4.5$ which yields $T_{\rm K} =  4$~K\@.\cite{Gruner1974} The value of $T_{\rm K}$ so obtained is close to the value of $T_{\rm K} =  5.6$ -- 6.9~K reported by Martin {\it et al}. based on the analysis of heat capacity data. \cite{Martin1996}

Figure~\ref{fig:MT-MH}(b) shows the heat capacity $C_{\rm p}$ versus $T$ for   CeCuGa$_{3}$ and its non-magnetic analog LaCuGa$_{3}$ measured in $H = 0$ T and $H = 2$~T\@. While no anomaly is seen in the $C_{\rm p}(T)$ of  LaCuGa$_{3}$, the $C_{\rm p}(T)$ of  CeCuGa$_{3}$ shows a well pronounced anomaly near 2.3~K associated with the magnetic phase transition as inferred from the $\chi(T)$ data above. Furthermore, it is seen that with the application of magnetic field, the $C_{\rm p}(T)$  peak position in CeCuGa$_{3}$ moves towards the higher temperature side. At $H= 2$~T the peak moves to 3.9~K from 2.3~K at $H=0$ T. Within the mean field theory such a behavior with magnetic field is generally considered to be an indication for ferromagnetic coupling which would contrast the deduction of dominant antiferromagnetic interaction inferred from the negative $\theta_{p}$. The broadening of heat capacity peak in magnetic field can be attributed to Zeeman splitting of CEF-split ground state doublet. The magnetic structure determined from the neutron diffraction measurement, as discussed in the next section, reveals an antiferromagnetic coupling in the $ab$-plane and a ferromagnetic coupling along $c$-direction. As such, we infer that at low fields the dominating exchange interaction is antiferromagnetic in nature whereas at high fields the dominating exchange interaction becomes ferromagnetic in nature. A close look of $M(H)$ curve presented in the upper inset of Fig.~\ref{fig:MT-MH}(a) suggests that at fields near 2~T where the nonlinearity and saturation tendency is prominent one should expect a dominating ferromagnetic exchange interaction.

The low-$T$ $C_{\rm p}(T)$ data of LaCuGa$_{3}$ are well described by $C_{\rm p}(T) = \gamma_{\rm n} T + \beta T^3$, allowing us to estimate $ \gamma$ and  $\beta$. A linear fit of $C_{\rm p}/T$ vs $T^2$ plot in the $T$-range 2~K to 7~K provides a Sommerfeld coefficient $ \gamma = 4.2(3)$~mJ/mol\,K$^2$ and $\beta = 0.28(2)$~mJ/mol\,K$^4$. We estimate the Debye temperature $\Theta_{\rm D} = (12 \pi^{4} R n/5 \beta )^{1/3}$  = 328(5)~K, where $R$ is the molar gas constant, and $n=5$ the number of atoms per formula units. Due to the presence of magnetic correlations at low-$T$, we fit the $C_{\rm p}/T$ versus $T^2$ for CeCuGa$_{3}$ in the $T$-range 10~K to 20~K which gives a Sommerfeld coefficient $ \gamma = 99(2)$~mJ/mol\,K$^2$ that is significantly enhanced compared to the $ \gamma $ value of LaCuGa$_{3}$. Martin {\it et al}. \cite{Martin1996} reported a value of $ \gamma = 150$~mJ/mol\,K$^2$ for antiferromagnetically ordered  CeCuGa$_{3}$ ($I4\,mm$) whereas Joshi {\it et al}. \cite{Joshi2012} found a value of $ \gamma = 20$~mJ/mol\,K$^2$ for ferromagnetically ordered CeCuGa$_{3}$ ($I4/mmm$).

The magnetic contributions to the heat capacity $C_{\rm mag}(T)$ and magnetic entropy $S_{\rm mag}(T)$ are shown in Fig.~\ref{fig:MT-MH}(c). The $S_{\rm mag}$ attains a value of 3.2~J/mol\,K at the transition temperature which is much lower than the value of $R \ln2$ expected for a doublet ground state, the $R \ln2$ value is attained near 10.3~K [see inset of Fig.~\ref{fig:MT-MH}(c)]. The  $S_{\rm mag}$  attains a value of $\frac{1}{2} R \ln2$ near 2.2~K, therefore the empirical relation for the Kondo temperature, $T_{\rm K} = 2 \, T (S_{\rm mag} = \frac{1}{2} \ln2)$ per Ce atom yields  $T_{\rm K} \approx 4.4$~K\@. This value of $T_{\rm K} $ is close to the value of $T_{\rm K} = 4$~K deduced above from $\theta_{\rm p}$. One can also estimate $T_{\rm K}$ from the jump in $C_{\rm mag}(T)$ at $T_{\rm N}$ using the universal plot of $\delta C_{\rm mag}$ versus $T_{\rm K}/T_{\rm N}$ for Ce- and Yb-based Kondo lattice systems.\cite{Besnus1992, Blanco1994} Our value of $\delta C_{\rm mag} = 3.5$~J/mol\,K [Fig.~\ref{fig:MT-MH}(c)] corresponds to $T_{\rm K}/T_{\rm N} \approx 1.7 $. Accordingly, for $T_{\rm N} = 2.3$~K we get $T_{\rm K} \approx 3.9$~K, which is again quite consistent with the values obtained from other estimates. From the quasielastic linewidth of low energy inelastic neutron scatttering data we obtain $T_{\rm K} \approx 6$~K (see Fig.~\ref{fig:INS-QENS}, Appendix).

The $C_{\rm mag}(T)$ presents a broad Schottky-type anomaly centered around 30~K [see Fig.~\ref{fig:MT-MH}(c)] which could be reproduced by a three-level crystal electric field scheme. Under the effect of crystal field the ($2J + 1$)-fold degenerate ground state multiplet of Kramers ion Ce$^{3+}$ ($J = 5/2$) splits into three doublets. From the analysis of $C_{\rm mag}(T)$ data by three-level CEF scheme \cite {Layek2009, Prasad2012, Anand2016} we find that the first excited doublet is situated at $\Delta_1 = 70$~K and the second excited doublet at $\Delta_2 = 240$~K\@. The fit of $C_{\rm mag}(T)$ by this CEF level scheme is shown by the solid red curve in Fig.~\ref{fig:MT-MH}(c). The $S_{\rm mag}(T)$  attains a value of $R \ln4$ near 66~K which further supports the deduced splitting energy between the ground state doublet and the first excited doublet to be 70~K\@. Our $C_{\rm mag}(T)$, $S_{\rm mag}(T)$ and CEF scheme of CeCuGa$_3$ is in agreement with that reported by Joshi {\it et al}.\ \cite{Joshi2012}. Joshi {\it et al}.\ reported a value of $S_{\rm mag}(T)$ close to $R \ln(6)$ at 250~K which is consistent with our $\Delta_2 = 240$~K\@.

\begin{figure}
\includegraphics[width=\columnwidth]{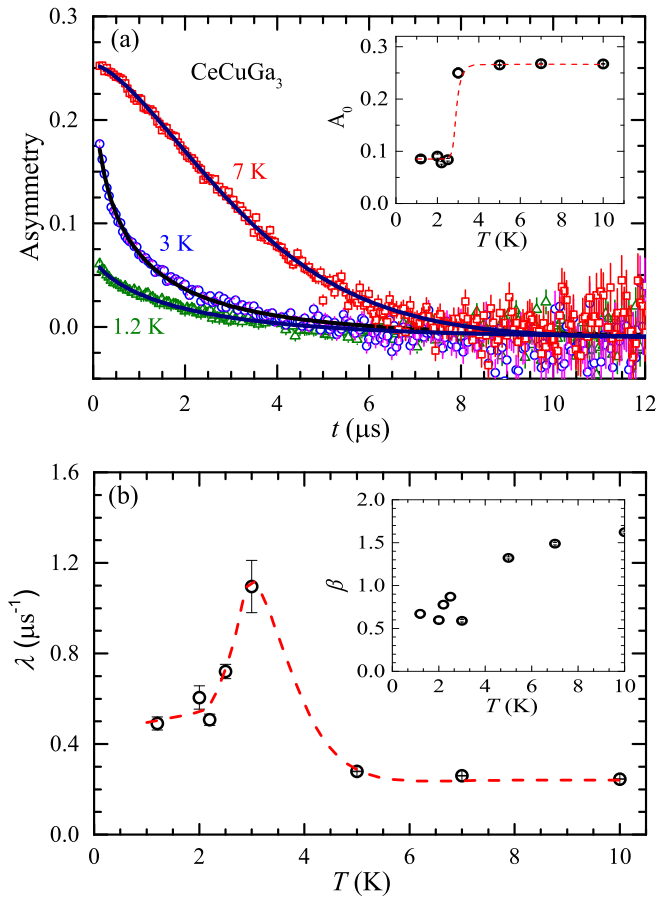}
\caption {(a) Zero-field (ZF) $\mu$SR asymmetry versus time $t$ spectra of CeCuGa$_{3}$ at three indicated representative temperatures. Solid curves are the fits of the ZF-$\mu$SR spectra by Eq.~(\ref{eq:exp}). Inset: Temperature $T$ depedence of the fit parameter muon initial asymmetry $A_0$. (b) $T$ dependence of relaxation rate $\lambda$. Inset: $T$ dependence of exponent $\beta$. Dashed lines are the guide to eye.}
\label{muSR}
\end{figure}

\section{\label{muSR} Muon Spin Relaxation}

Magnetic ordering in CeCuGa$_{3}$  was further probed by muon spin relaxation measurements in zero-field (ZF). The representative ZF-$\mu$SR spectra collected at 1.2~K, 3~K, and 7~K are shown in Fig.~\ref{muSR}(a). As can be seen from Fig.~\ref{muSR}(a), a loss in initial asymmetry clearly depicts a magnetic phase transition in CeCuGa$_{3}$. The  ZF-$\mu$SR data are well described by a stretched exponential function, i.e.,
\begin{equation}
A = A_0 \exp[-({\lambda}t)^\beta]+A_{\rm BG},
\label{eq:exp}
\end{equation} 
where $A_0$ is the initial asymmetry, $\lambda$ is the muon spin relaxation rate, $\beta$ is an exponent, and $A_{\rm BG}$ is a constant accounting for the contribution from the sample holder. The value of $A_{\rm BG} = 0.0106$ was determined by fitting the $\mu$SR data at 10~K, and kept fixed to this value while fitting the data at other temperatures. The fits of the  representative ZF-$\mu$SR spectra by Eq.~(\ref{eq:exp}) are shown as solid curves in Fig.~\ref{muSR}(a). The fitting parameters $A_0$, $\lambda$ and $\beta$ obtained from the fits are shown in Fig.~\ref{muSR}.  A huge loss in initial asymmetry $A_0(T)$ [inset of Fig.~\ref{muSR}(a)], and a peak in $\lambda(T)$ [Fig.~\ref{muSR}(b)] clearly mark a magnetic phase transition near 3~K\@. Furthermore, a significant decrease in exponent $\beta$ is found to accompany the magnetic phase transition. The loss in $A_0$ in the ordered state reflects a highly damped $\mu$SR signal in the ordered state, which can be associated with the incommensurate nature of magnetic structure as inferred from the neutron diffraction data (discussed below). For an incommensurate antiferromagnetic order there will be a broad distribution of local fields at the muon stopping sites, resulting in a stronger damping of $\mu$SR signal below $T_{\rm N}$ (relaxation rate depends on the width of the field distribution).

\begin{figure} 
\includegraphics[width=\columnwidth]{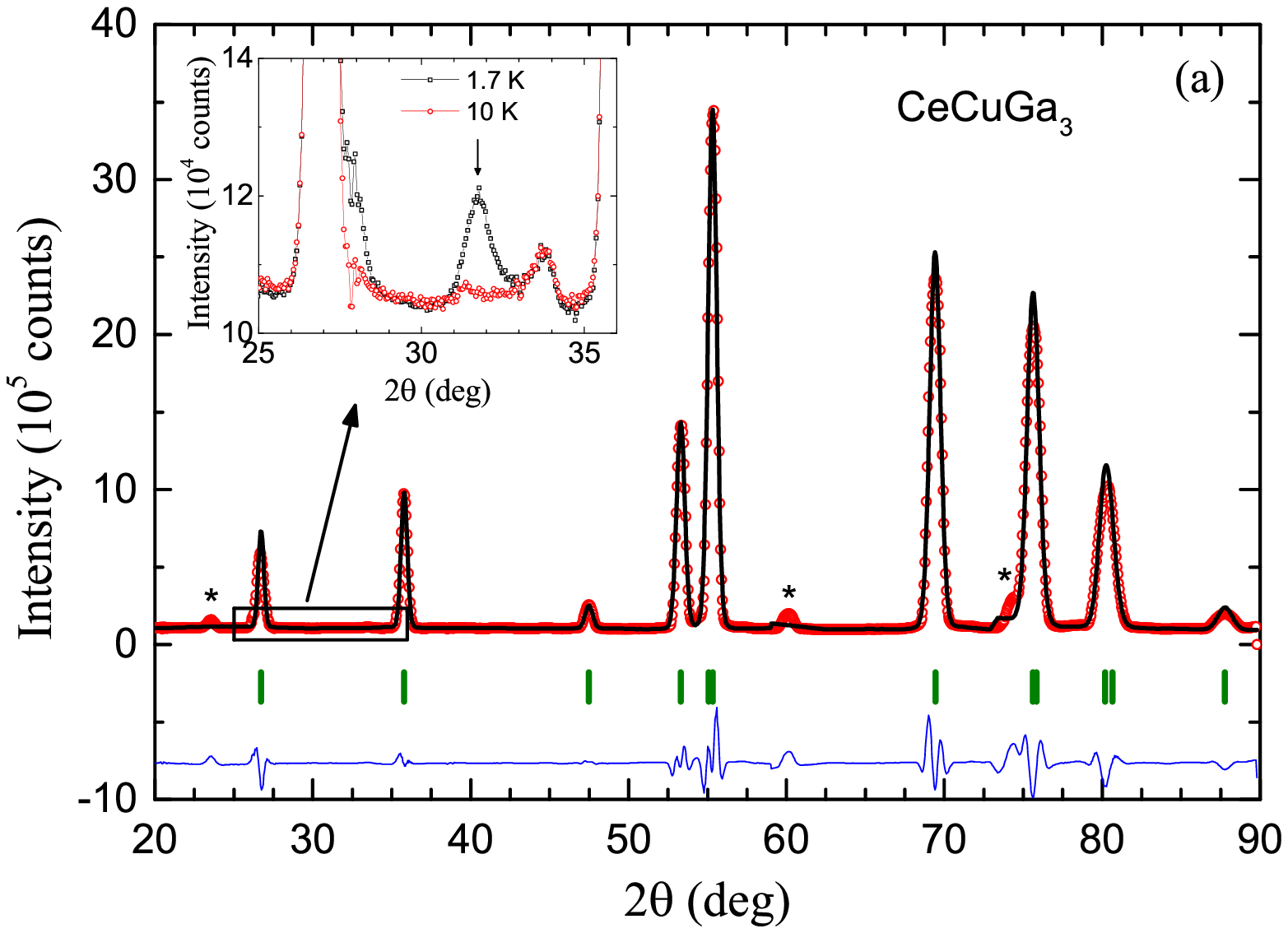}\vspace{0.3cm}
\includegraphics[width=\columnwidth]{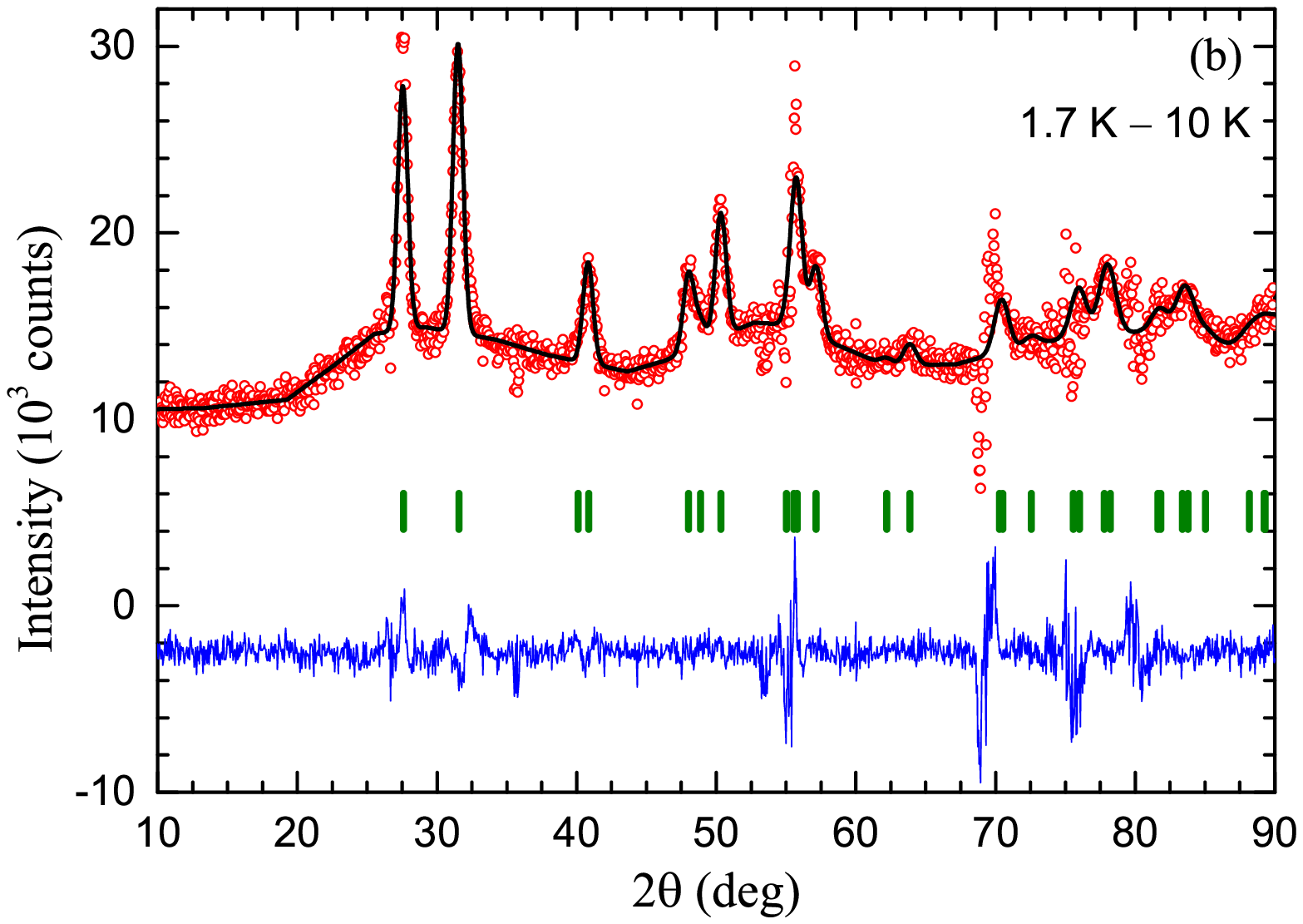}
\caption{\label{fig:ND} (a) Neutron diffraction (ND) pattern of CeCuGa$_{3}$ recorded at 10~K along with the structural refinement profile\@. Asterisks mark the extrinsic peaks. Inset: expanded view of ND pattern between 25$^\circ$ and 36$^\circ$, and a comparison of ND patterns at 1.7~K (ordered state) and 10~K (paramagnetic state) to highlight the presence of magnetic Bragg peaks. (b) Magnetic diffraction pattern at 1.7~K (after subtracting the 10~K nuclear pattern) together with the calculated magnetic refinement pattern. The difference between the experimental and calculated intensities is shown by the blue curve at the bottom.}
\end{figure}

\begin{figure} 
\includegraphics[width=3in]{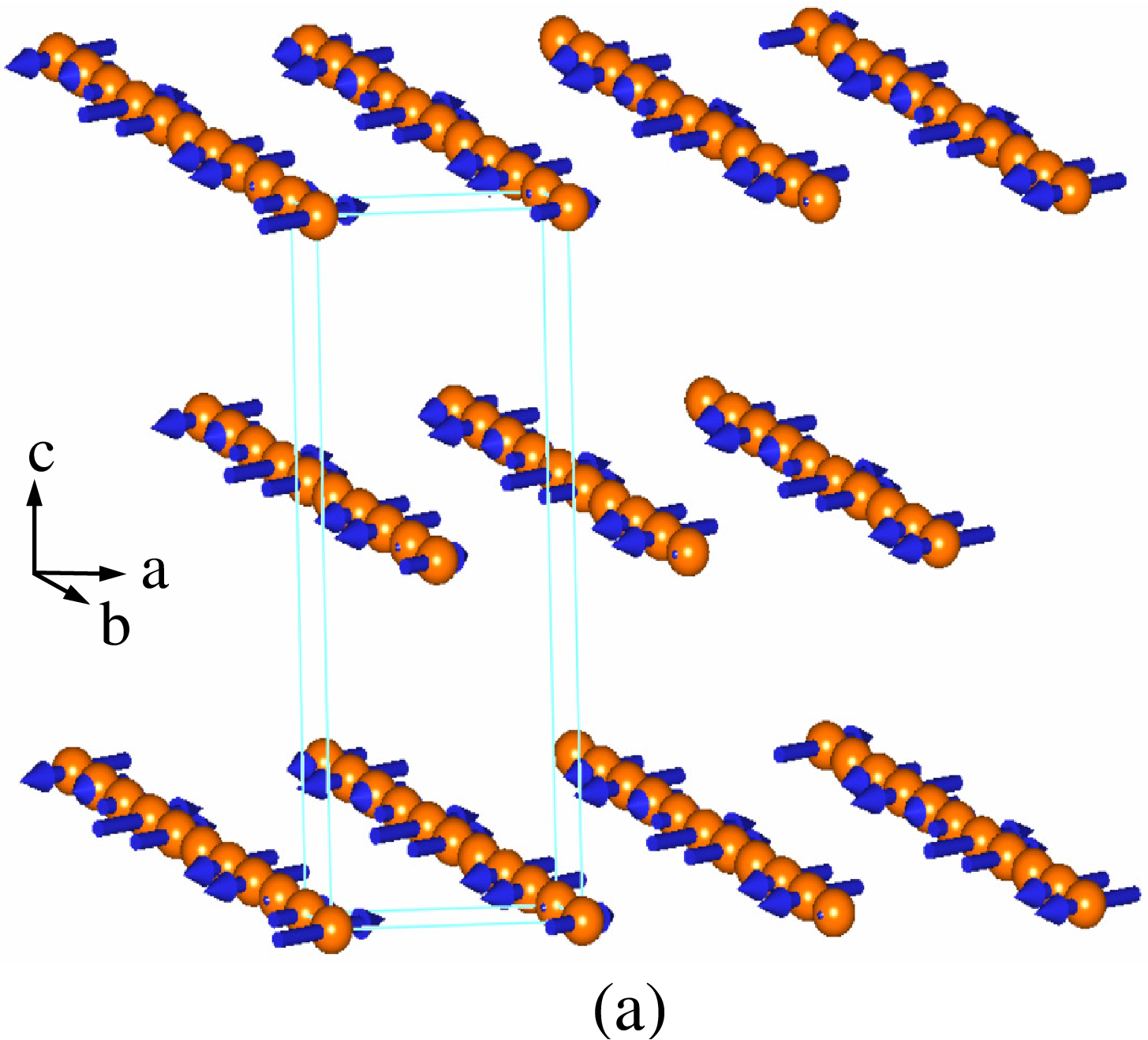}\vspace{0.4cm}
\includegraphics[width=\columnwidth]{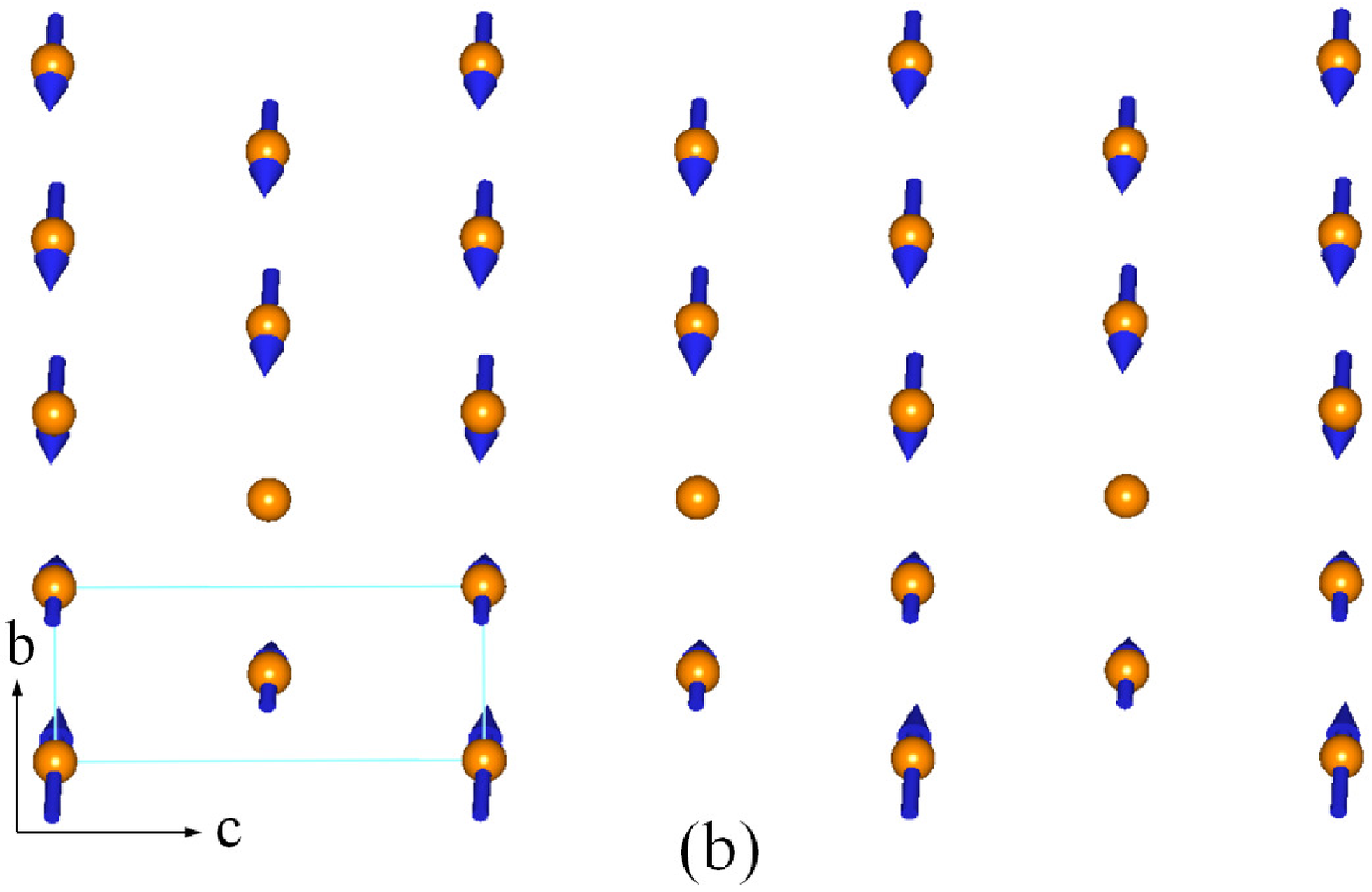}
\caption{\label{fig:Mag_struct} (a) Incommensurate magnetic structure (longitudinal spin density wave) of CeCuGa$_{3}$ obtained from the refinement of magnetic diffraction pattern at 1.7~K presented in Fig.~\ref{fig:ND}(b). (b) A projection of magnetic structure onto the $bc$ plane showing ferromagnetic arrangement of moments along the $c$-direction.  }
\end{figure}

\section{\label{ND} Neutron Powder Diffraction}

In order to determine the nature of the long range magnetic ordering in CeCuGa$_{3}$ we also performed neutron diffraction measurements on powdered sample at 1.7~K (ordered state) and 10~K (paramagnetic state). The neutron powder diffraction pattern collected at 10~K and the calculated pattern for the nuclear structure are shown in Fig.~\ref{fig:ND}(a). The structural Rietveld refinement for the BaNiSn$_{3}$-type tetragonal (space group $I4\, mm$) structure yielded the lattice parameters  $a=4.2322(3)$~{\AA} and $c=10.4284(6)$~{\AA} which are in very good agreement with the literature values.\cite{Martin1998} We also see a few weak unindexed peaks (marked with asterisks in Fig.~\ref{fig:ND}(a)) which could be due to the presence of tiny amount of unidentified impurities in sample.

A comparison of the ND patterns at  1.7~K (ordered state) and 10~K (paramagnetic state) clearly displays the appearance of magnetic Bragg peaks at 1.7~K\@. The inset of Fig.~\ref{fig:ND}(a) shows the expanded view of the two ND patterns over the $2\theta$ range of 25$^\circ$ to 36$^\circ$. The difference plot of 1.7~K and 10~K ND patterns presented in Fig.~\ref{fig:ND}(b) clearly  shows the presence of a number of magnetic Bragg peaks at 1.7~K\@.  All the magnetic Bragg peaks are well indexed by a propagation wave vector ${\bf k} = (0.148, 0.148, 0)$ implying an incommensurate magnetic structure. Our {\bf k} is very similar to the one reported by Martin {\it et al}. \cite{Martin1998} who found ${\bf k}=(0.176, 0.176, 0)$ to index the magnetic Bragg peaks in the case of polycrystalline sample,  and the same magnetic propagation vector ${\bf k}=(0.176,0.176,0) $ was also used to index the magnetic Bragg peaks of single crystal CeCuGa$_{3}$. 

\begin{figure} 
\includegraphics[width=2.5in]{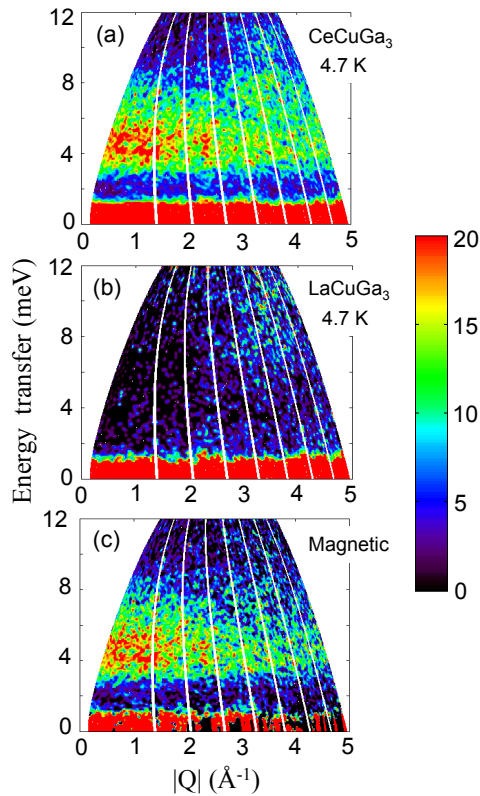}
\caption{\label{fig:INS-contour} Inelastic neutron scattering response, a color-coded contour map of the intensity (in unit of mb/meV/sr/f.u.), energy transfer $E$ versus momentum transfer $Q$ for (a) CeCuGa$_{3}$, and (b) LaCuGa$_{3}$ measured at 4.7~K with an incident energy $E_{i} = 15$~meV. (c) Magnetic contribution to CeCuGa$_{3}$ estimated after subtracting the phonon contribution using the data of LaCuGa$_{3}$ at 4.7 K.}
\end{figure}

Magnetic symmetry analysis using the program BASIREPS \cite{Rodriguez, Ritter2011} for {\bf k}~$ = (0.148, 0.148, 0)$ and space group $I4\,mm$ revealed two allowed magnetic representations for Ce atoms occupying the Wyckoff position $2a$. The first one has only one basis vector of type  (1 $-1$ 0) corresponding to magnetic moments lying in the tetragonal basal plane, while the second one has two basis vectors of type  (1 1 0) and (0 0 1). Only the second magnetic representation is able to reproduce the magnetic intensity present in the diffraction pattern at 1.7 K with apparently no contribution from the basis vector (0 0 1), and all Ce moments lying in the basal plane corresponding to the basis vector (1 1 0).

The refinement of the difference data set ${\rm 1.7~K - 10~K}$ which contains only the magnetic diffraction intensity is shown in Fig.~\ref{fig:ND}(b) and the corresponding magnetic structure is shown in Fig.~\ref{fig:Mag_struct}. The incommensurate magnetic structure corresponds to a longitudinal spin density wave where the magnetic moments point along the (1 1 0) direction. The maximum value of the ordered moment is found to be $m = 0.95(1)\,\mu_{\rm B}$/Ce. Our analysis of ND data reveals a different magnetic structure compared to the one proposed by Martin {\it et al}. \cite{Martin1998} who also reported an incommensurate magnetic structure but with a helical arrangement of moments having a spiral axis oriented at $35(2)^\circ$ from $c$-axis, which however does not represent one of the symmetry allowed solutions.

A view of the magnetic structure projected onto the $bc$ plane i.e., the arrangement of magnetic moments when viewed along the crystallographic $a$-direction is shown in Fig.~\ref{fig:Mag_struct}. It is seen that Ce moments are coupled ferromagnetically along the $c$-direction. This explains the indication of ferromagnetic correlation in the heat capacity measurement under the applied field discussed above in Sec.~\ref{MT-HC}.

\begin{figure}
\includegraphics[width=\columnwidth]{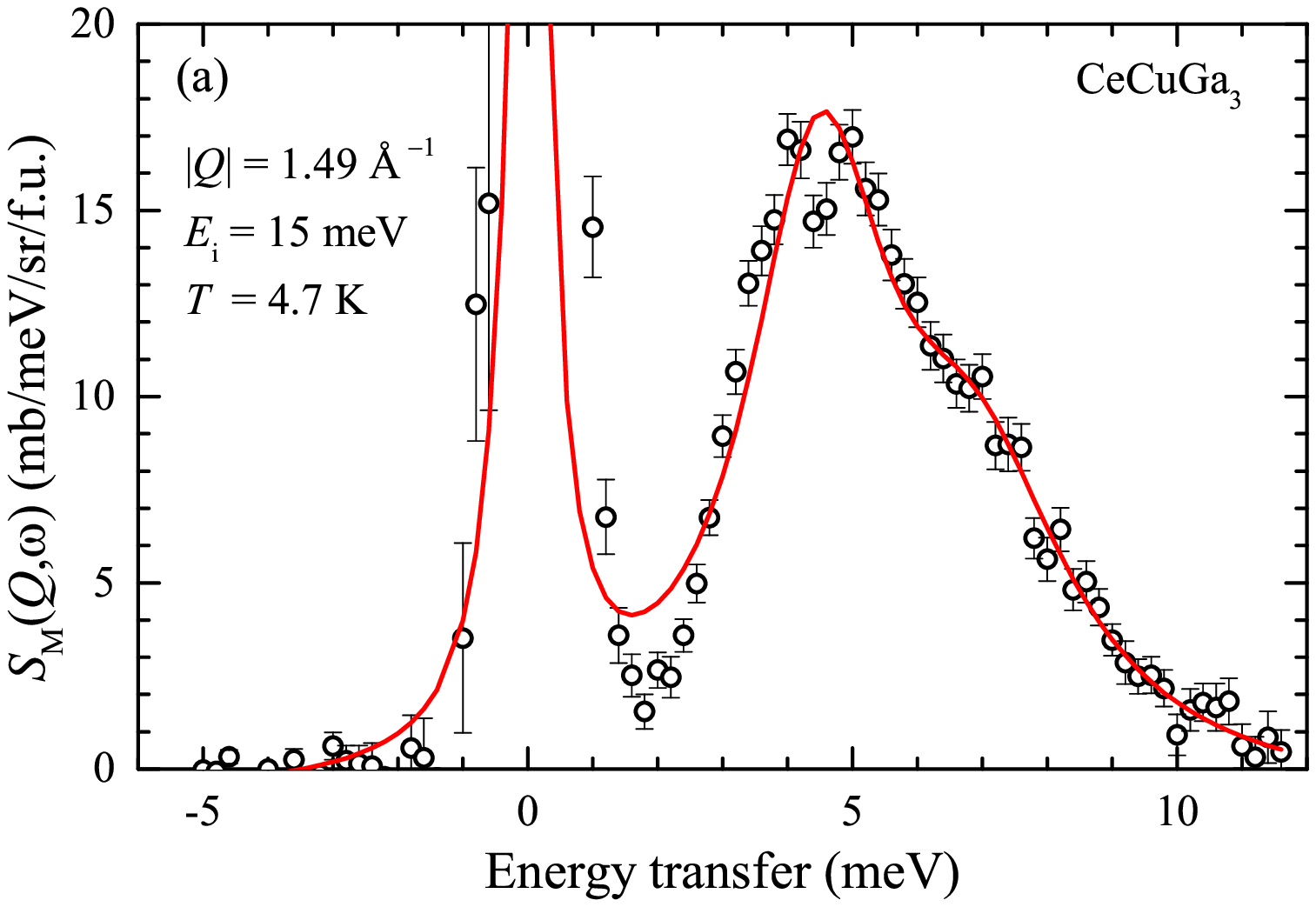}\vspace{0.05cm}
\includegraphics[width=\columnwidth]{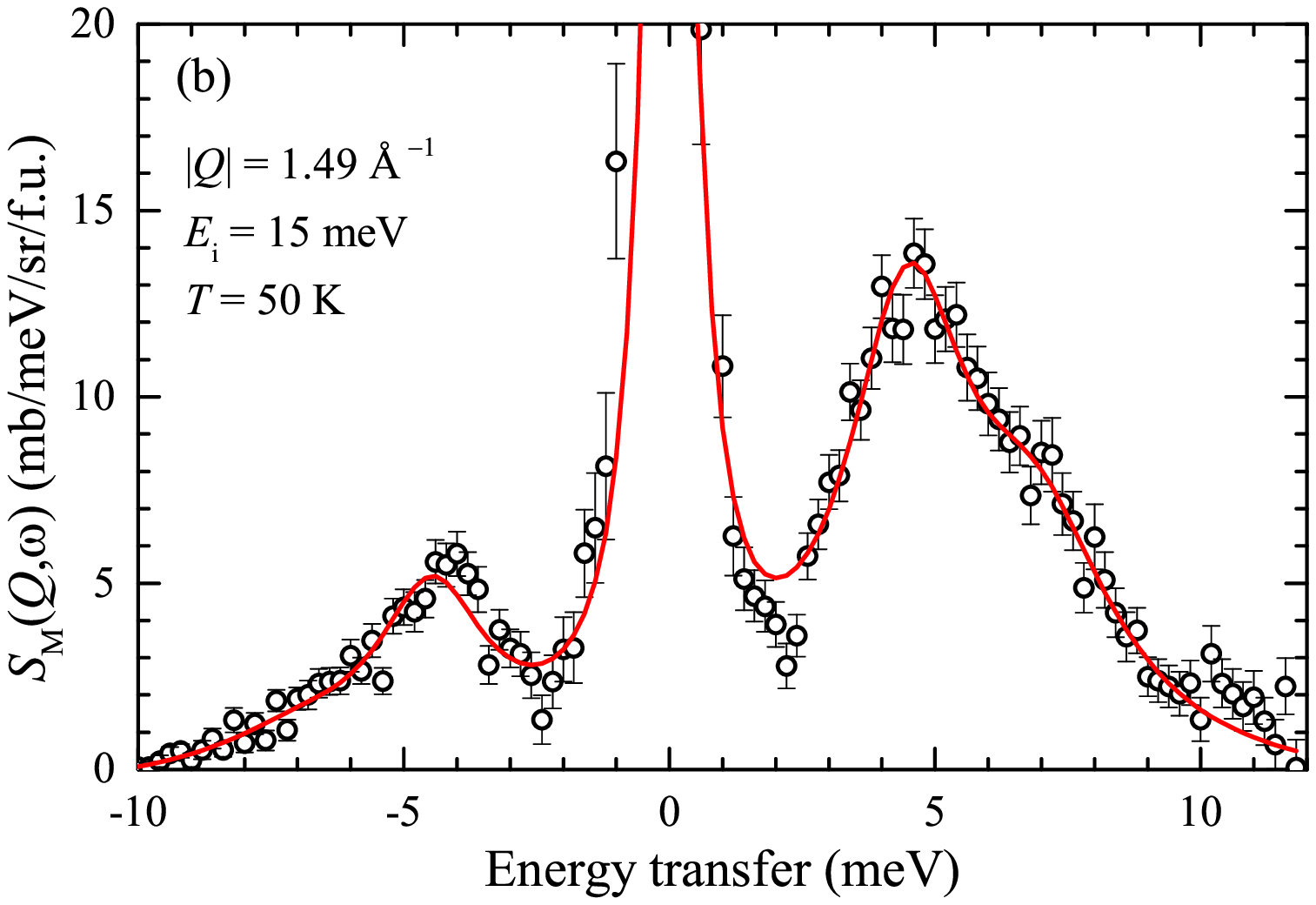}\vspace{0.05cm}
\includegraphics[width=\columnwidth]{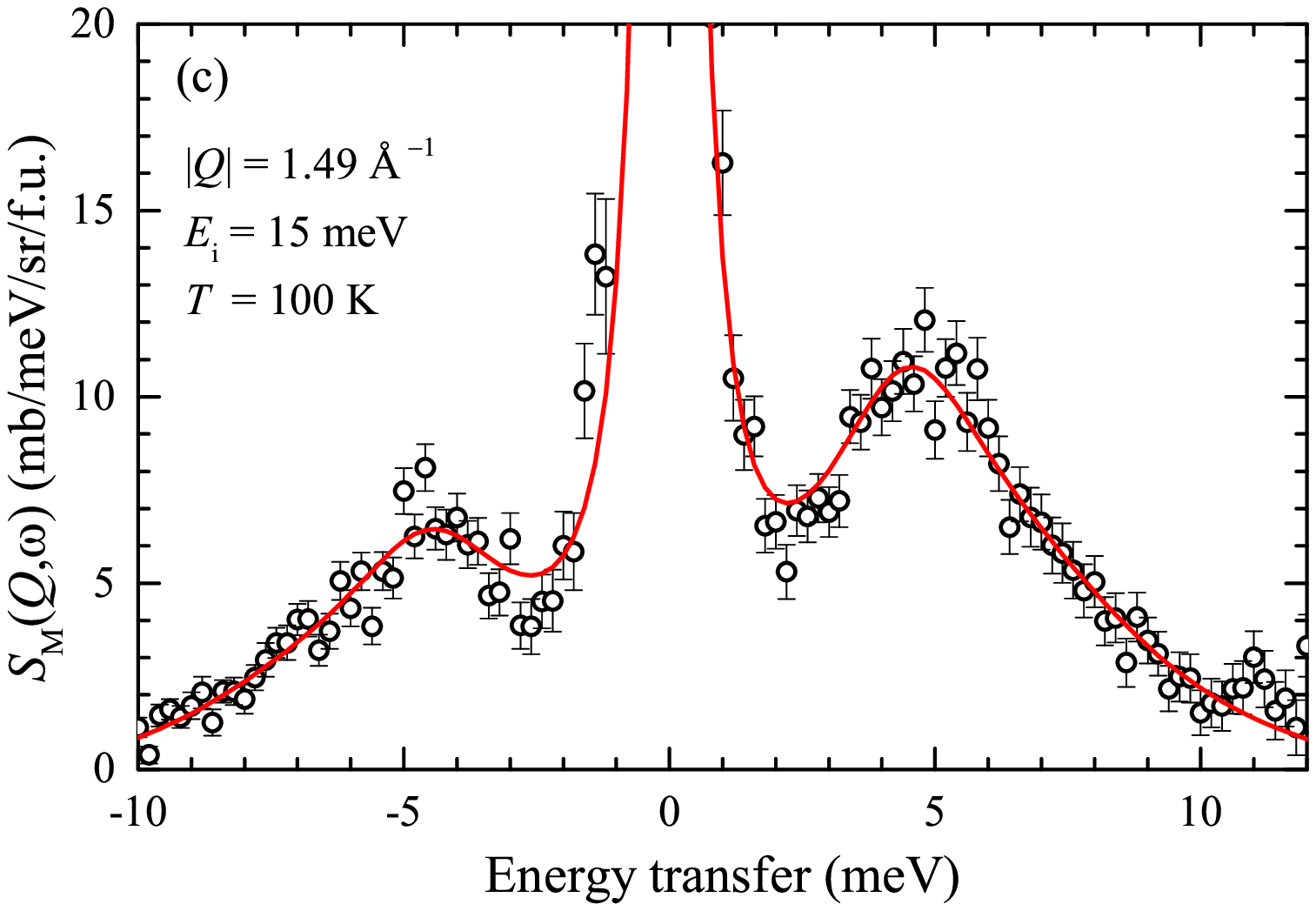}
\caption{\label{fig:INS-Smag1} $Q$-integrated ($0 \leq Q \leq 2.5$~{\AA}$^{-1}$) inelastic magnetic scattering intensity $S_{\rm M}(Q,\omega)$ versus energy transfer $E$ for CeCuGa$_{3}$ at $|Q|=1.49$~{\AA}$^{-1}$ measured with $E_{i}= 15$~meV at (a) 4.7~K, (b) 50~K, and (c) 100~K\@. The solid red curves are the fits of the data based on crystal electric field model [Eq.~(\ref{H-CEF}]].}
\end{figure}

\begin{figure} 
\includegraphics[width=\columnwidth]{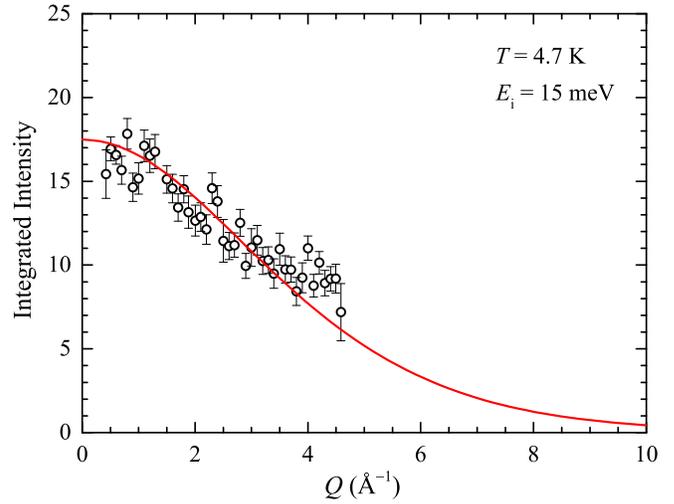}
\caption{\label{fig:formfactor} The $Q$ dependence of total intensity
integrated between 3.5 to 6~meV at 4.7~K for incident energy $E_i = 15$~meV. The solid line represents the square of the Ce$^{3+}$ magnetic form factor, scaled to 17.5 at $Q = 0$.}
\end{figure}

\section{\label{INS} Inelastic neutron scattering}

The color coded intensity maps showing inelastic neutron scattering responses from CeCuGa$_{3}$ and LaCuGa$_{3}$ measured with $E_{i} = 15$~meV at $T = 4.7$~K are shown in Fig.~\ref{fig:INS-contour}. While only a phononic excitation is seen for LaCuGa$_{3}$, at low-$Q$ a magnetic excitation, near 4.5~meV, is quite evident for CeCuGa$_{3}$. 
We see that the magnetic excitation in CeCuGa$_{3}$ is relatively broader reflecting the possibility of two closely situated excitations [see Fig.~\ref{fig:INS-contour}(c)]. In addition, the magnetic excitation energy does not change with increasing temperature from 4.7~K to 100~K which suggests that this excitation has its origin in the crystal electric field effect. Moreover, very weak and broad excitations were observed over 20~meV to 30~meV range in our INS measurements with $E_{i} = 40$~meV at 5~K and 100~K (see Figs.~\ref{fig:contour-40meV} and \ref{fig:INS-40meV} in Appendix).

\begin{figure} 
\includegraphics[width=8cm]{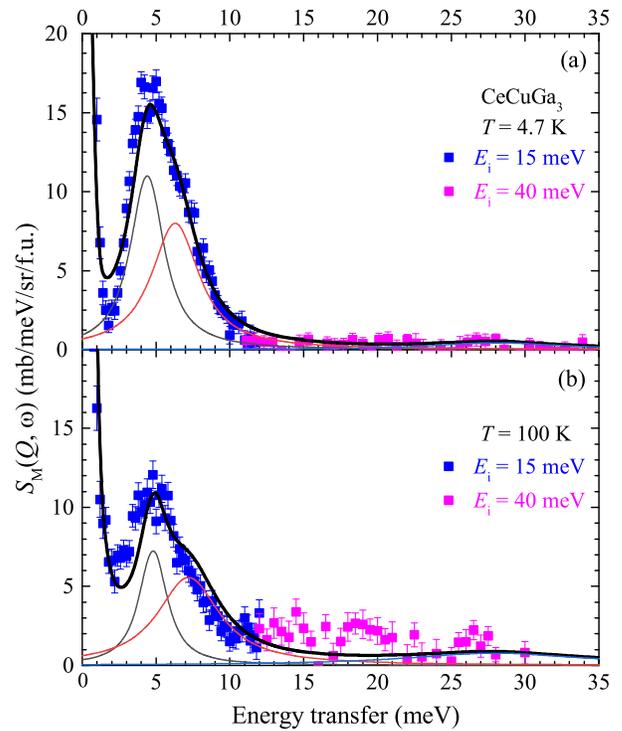}
\caption{\label{fig:Fig8CEFPhon} Magnetic scattering $S_{\rm M}(Q,\omega)$ vs energy transfer $E$ for CeCuGa$_3$ from the 15 meV (blue squares, below 12 meV) and 40 meV data (magenta squares, above 12 meV) at (a) 4.7~K, and (b) 100~K\@. The thick solid black lines represent the fit based on CEF-phonon model using Eq.~(\ref{H-CEF-ph}). The thin lines correspond to the contributions from the three excitations.}
\end{figure}

\begin{figure*} 
\includegraphics[width=17cm]{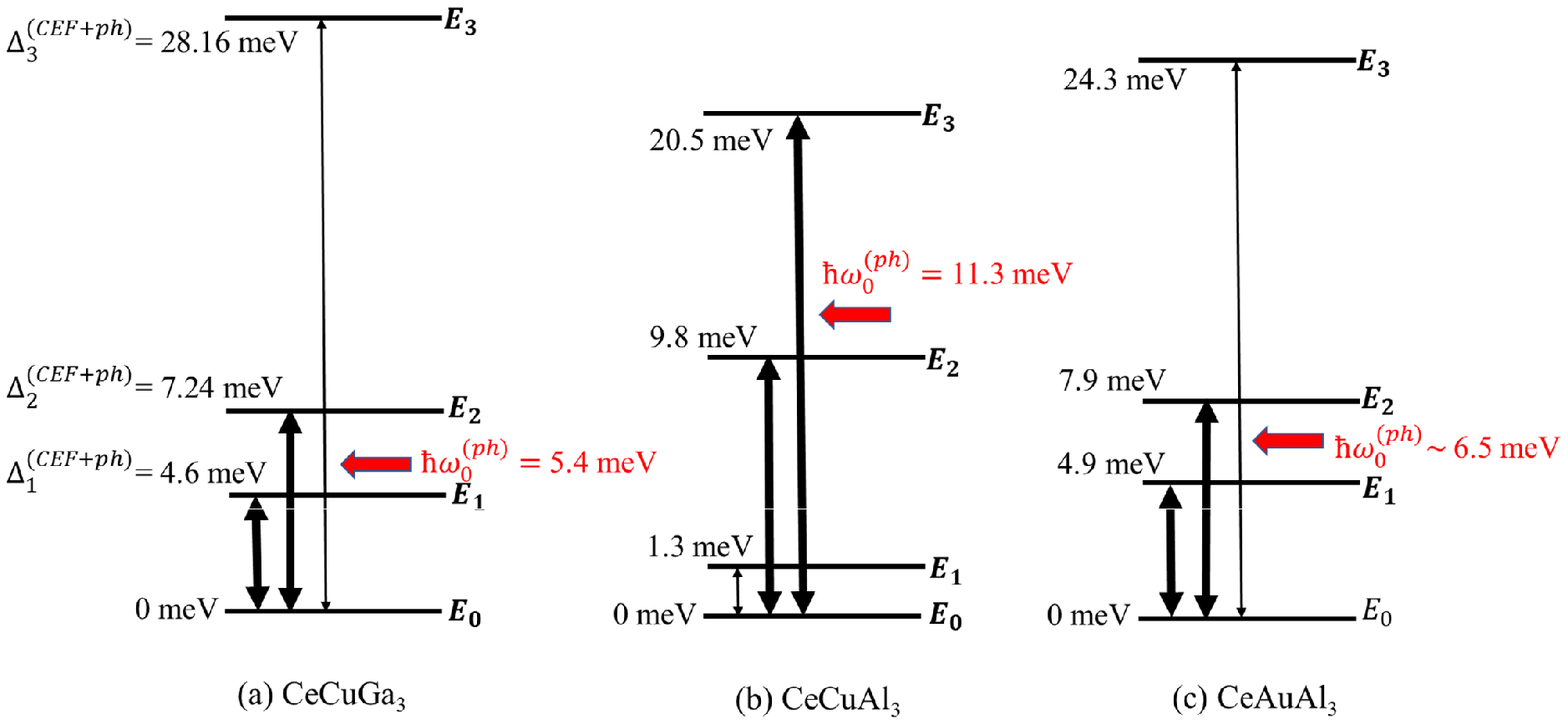}
\caption{\label{fig:CEF-phonon} (a) The schematic energy levels of CEF-phonon coupled model along with the position of the phonon energy (red arrow) obtained from the fitting of INS data of CeCuGa$_{3}$. Vertical arrows represent the excitations from the ground state. The magneto-elastic excitations are shown by thick arrows. The CEF-phonon schematic is compared with those of (b) CeCuAl$_{3}$ \cite{Adroja2012} and (c) CeAuAl$_{3}$.\cite{Petr2019}} 
\end{figure*}

The magnetic excitations are more clear in $Q$-integrated one-dimensional energy cuts as shown in Fig.~\ref{fig:INS-Smag1}. The magnetic contribution $S_{\rm M}(Q, \omega)$ to INS response was separated out by subtracting off the phononic contribution using the INS  response of LaCuGa$_{3}$. The $S_{\rm M}(Q, \omega) = S(Q, \omega)_{\rm CeCuGa_{3}} - \alpha S(Q, \omega)_{\rm LaCuGa_{3}}$ with $\alpha = 0.82$, the ratio of neutron scattering cross sections of CeCuGa$_{3}$ and LaCuGa$_{3}$. From the one-dimensional energy cuts of E$_i$= 15 meV in Fig.~\ref{fig:INS-Smag1}, it is seen that the CEF excitation in CeCuGa$_{3}$ consists of two closely situated excitations which is better seen in our fit of the INS data by a model based on CEF (discussed later). 

Figure~\ref{fig:formfactor} shows the $Q$-dependent integrated intensity integrated over an energy range of 3.5 to 6~meV for INS data collected with $E_{i} = 15$~meV at 4.7~K\@. The solid red curve is the theoretical value of the square of the Ce$^{3+}$ magnetic form factor $[F^2(Q)]$ \cite{Brown}, which is scaled to 17.5 times at $Q = 0$. The fact that the integrated intensity is consistent with $[F^2(Q)]$ implies that the inelastic excitations mainly result from the single-ion CEF transitions.

The CEF Hamiltonian for the tetragonal symmetry (point symmetry $C_{4v}$) of the Ce$^{3+}$ ions is given by
\begin{equation}\label{H-CEF}
 H_{\rm CEF} = B_{2}^{0}O_{2}^{0} + B_{4}^{0}O_{4}^{0} + B_{4}^{4}O_{4}^{4}
\end{equation}
where $B_{n}^{m}$ are CEF parameters and $O_{n}^{m}$ are the Stevens operators.\cite{Stevens} $B_{n}^{m}$ parameters need to be estimated by fitting the experimental data, such as single crystal susceptibility and/or inelastic neutron scattering data.  For the analysis of INS data, we use a Lorentzian line shape for both quasi-elastic (QE) and inelastic excitations.

In order to obtain a set of CEF parameters that consistently fit the INS data at different temperatures, we performed a  simultaneous  fit  of  INS  data  at  4.7~K,  50~K  and  100~K. The CEF parameters obtained from the simultaneous fits are (in meV) $B_2^0$ = 0.195, $B_4^0$ = 0.0187 and $B_4^4$ = 0.0190. The solid curves in Fig.~\ref{fig:INS-Smag1}(a)-(c) show the fits of the INS data. We find that the splitting energy between the ground state doublet and the first excited doublet is $\Delta_1 = 4.5$~meV (52~K) and that between the ground state and the second excited doublet is $\Delta_2 = 6.9$~meV (80~K). Although the fits to INS data look very good, the value of the overall CEF splitting of 6.9 meV is much smaller than the value expected based on the magnetic entropy and heat capacity analysis, which indicates overall CEF splitting about 240-250 K (21-22 meV). Furthermore, this set of the CEF parameters fails to explain the observed magnetic susceptibility of single crystal as well as polycrystal CeCuGa$_3$. 

\begin{figure} 
\includegraphics[width=\columnwidth]{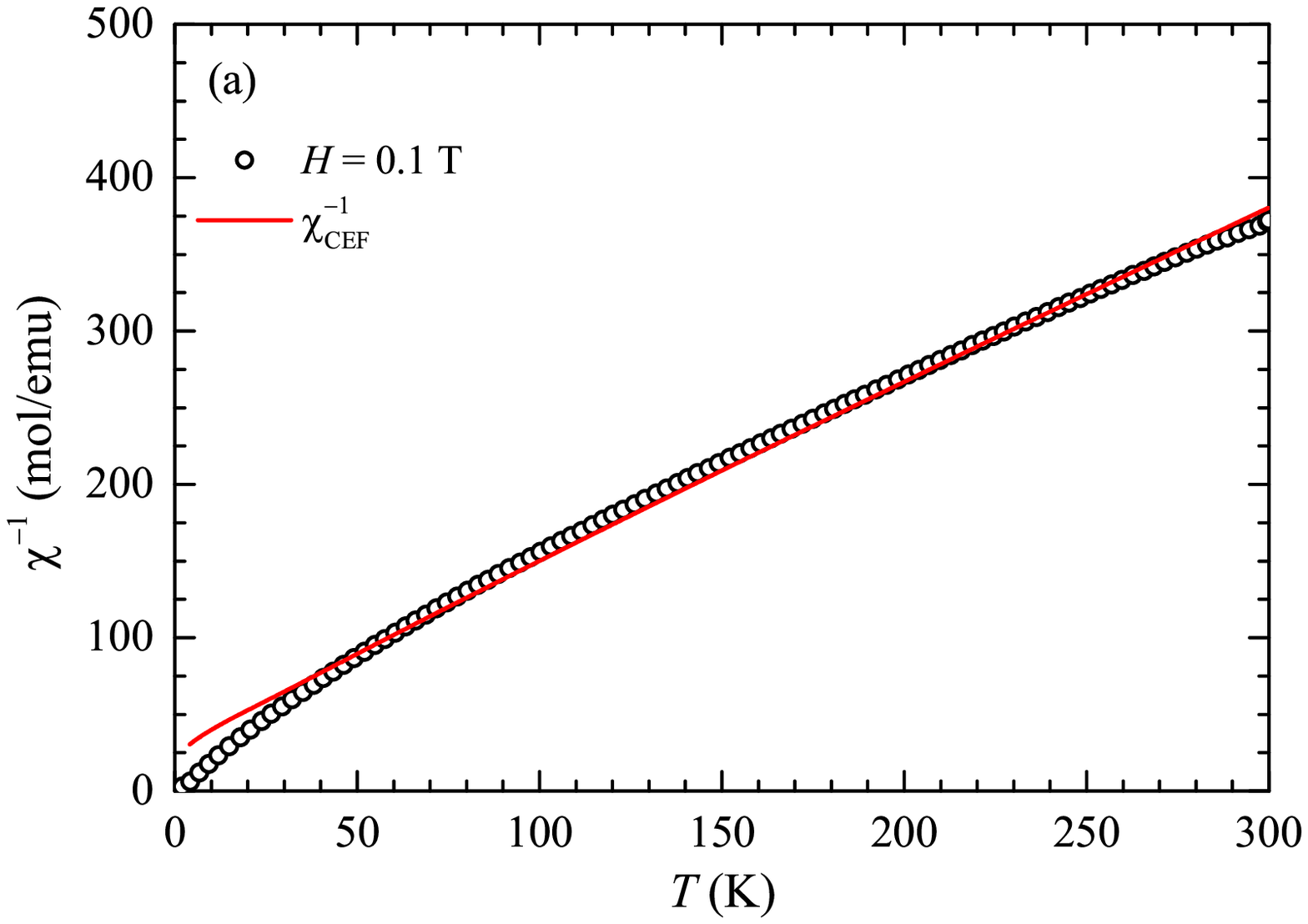}\vspace{0.1cm}
\includegraphics[width=\columnwidth]{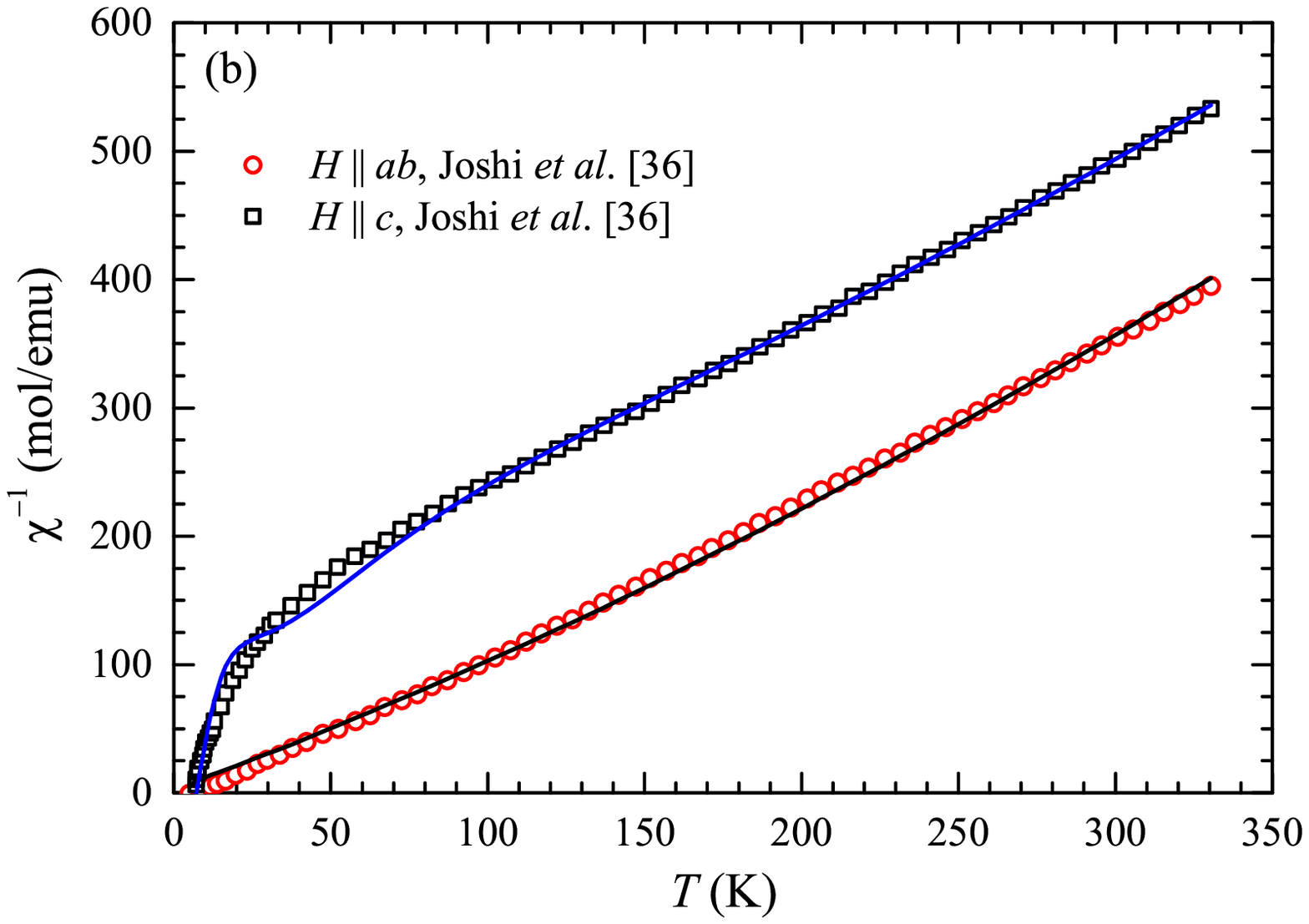}\vspace{0.1cm}
\caption{\label{fig:Chi_CEF} (a) A comparison of magnetic susceptibility $\chi(T)$ of polycrystalline CeCuGa$_{3}$ and CEF susceptibility $\chi_{\rm CEF}(T)$ (obtained from the INS fit) plotted as inverse susceptibility measured in an applied field $H =0.1$~T\@. The solid red line represents the CEF susceptibility. (b) A comparison of CEF susceptibility (solid lines) with the susceptibility of single crystal CeCuGa$_{3}$ taken from Ref.~\cite{Joshi2012}. The $\chi_{\rm CEF}$ corresponds to the CEF parameters obtained from the fitting of the INS data by CEF-phonon model [Eq.~(\ref{H-CEF-ph})].}
\end{figure}

Joshi {\it et al}. \cite{Joshi2012} determined the CEF parameters for ferromagnetically ordered CeCuGa$_{3}$ ($I4/mmm$) using the single crystal magnetic susceptibility data. They also obtained a positive value for CEF parameter $B_{2}^{0}$ which is consistent with the $ab$-plane as the easy plane of magnetization. However, the values of their $B_{n}^{m}$  parameters: $B_{2}^{0}=11.0$~K (0.948~meV), $B_{4}^{0}=0.127$~K (0.011~meV) and  $B_{4}^{4}= -3.0$~K ($-0.259$~meV) differ significantly from ours. They found the first excited state at 4.3 meV (50 K) which is in good agreement with ours. Nevertheless, they found the second excited state at 19.6~meV (228~K) which is drastically different from our excitations (4.5 and 6.9~meV). We used Joshi {\it et al}. CEF parameters and simulated INS spectra at 4.7 K and found excitaions at 0.44~meV (not at 4.3~meV) and 19.7~meV. Furthermore, Joshi {\it et al}. \cite{Joshi2012} also reported that the magnetic entropy $R \ln6$ is attained near 250~K (21.5~meV), which also suggests a much higher overall splitting energy compared to 6.9~meV. Our analysis of $C_{\rm mag}(T)$ in Sec.~\ref{MT-HC} also suggests an overall splitting of 240~K (20.7~meV).  CeRhGe$_3$ and CeIrGe$_3$, which are isostructural to CeCuGa$_3$, also have higher overall splitting energy. The INS measurements reveal two well-defined CEF excitations at 7.5 and 18 meV in CeRhGe$_3$ \cite{Hillier2012}, and  at 9.7 and 20.9 meV in CeIrGe$_3$ \cite{Anand2018}. 

All these suggest that the two magnetic excitations (4.5~meV and 6.9~meV) detected in the INS spectra of CeCuGa$_3$ do not originate only from single-ion CEF transition. Therefore, it is very likely that the first CEF doublet at 70~K (6~meV), as deduced from the analysis of $C_{\rm mag}(T)$, splits into two levels due to magneto-elastic CEF-phonon coupling, resulting in two magnetic excitations near 4.5~meV and 6.9~meV. The presence of an additional excitation on account of CEF-phonon coupling has been seen in related compounds CeCuAl$_{3}$  \cite{Adroja2012} and CeAuAl$_{3}$.\cite{Petr2019} In the case of CeCuAl$_{3}$ INS revealed three magnetic excitations near 1.3, 9.8, and 20.5~meV. \cite{Adroja2012} According to Kramer’s degeneracy theorem for Ce$^{3+}$ ($J = 5/2$)  only two CEF excitations are expected, and the additional excitation in CeCuAl$_{3}$ arises from CEF-phonon coupling, which gives two excitations at 9.8 meV and 20.5 meV (referred as vibrons) due to coupling with the phonon modes near 14 meV.\cite{Adroja2012} On the other hand, INS study on single crystal CeAuAl$_{3}$ has revealed CEF-phonon excitations near 4.9 and 7.9~meV.\cite{Petr2019}

In order to consider the possibility of CEF-phonon coupling, and to understand the discrepancy between the CEF parameters obtained from the INS fit of 15 meV (at 4.7 K, 50 K, and 100 K) and that obtained from the single crystal susceptibility by Joshi {\it et al}. \cite{Joshi2012}, we have analyzed the 15 meV and 40 meV data (combining two sets of the data) at 4.7 K and 100 K, including the magnetoelastic term (MEL, i.e. CEF-phonon coupling term) to the CEF Hamiltonian given in Eq.~(\ref{H-CEF}). Hence, the total Hamiltonian is given as follow:
\begin{equation}\label{H-CEF-ph}
H_{\rm total}= H_{\rm CEF} +\hslash\omega_{0}(a_{u}^{+}a_u+1/2)+M^{\gamma}(a_u+a_{u}^{+})O_{u}
\end{equation}
where the first term is the tetragonal CEF Hamiltonian mentioned above [Eq.~(\ref{H-CEF})], the second term is the phonon Hamiltonian ($H_{\rm ph}$), and the third term is the (CEF-phonon)-coupling term ($H_{\rm CEF-ph}$). Here $\hbar \omega_{0}$ denotes the phonon energy, and $a_u^+$
 or $a_u$ are phonon creation or annihilation operators. $M^{\gamma}$ is the coupling parameter between CEF and phonon excitations and ($a_u+a_{u}^{+})O_{u}$ is the magnetoelastic CEF-phonon operator within orthorombic symmetry, $\gamma$-mode, the same as used in CeCu$_{x}$Ag$_{1-x}$Al$_3$ ($0.2 \leq x \leq 1$)\cite{Adroja2012,Fuente2021}, with $O_{u}=O_{2}^{2}=J_{x}^2-J_{y}^2=(J_{+}^2 + J_{-}^2)/2$.
 
 Figure~\ref{fig:Fig8CEFPhon} shows the fit (the solid black curves) to the combined 15 and 40~meV data at 4.7~K and 100~K according to Eq.~(\ref{H-CEF-ph}), i.e., including the phonon and MEL coupling terms to the CEF-Hamiltonian. This model explains very well the observed excitations in CeCuGa$_3$. Figure~\ref{fig:CEF-phonon}(a) shows the energy level diagram of CEF-phonon coupled model.  The value of fits parameters obtained are: $B_{2}^{0}=1.587(5)$~meV, $B_{4}^{0}=0.0143(2)$~meV and  $B_{4}^{4}=0.000(1)$~meV, $\hbar\omega_{0}=5.4(5)$~meV and  $M^{\gamma}= 0.15(05)$~meV.  The value of $\hbar\omega_{0}$ is slightly smaller than the phonon peak at 8~meV in LaCuGa$_3$, which was also found in CeCuAl$_3$ \cite{Adroja2012}. Moreover, $M^{\gamma}$ in CeCuGa$_{3}$ is approximately half of the value found in the CeCuAl$_{3}$.\cite{Adroja2012} Since $M^{\gamma}$ is proportional to the effective intrinsic magnetoelastic parameter, but inversely proportional to the phonon energy,\cite{Fuente2021} the reduction of CEF-phonon coupling intensity for CeCuGa$_{3}$ can only be due to an important decay of the intrinsic magnetoelastic interaction, assuming they have a similar elastic modules. It is also interesting to observe that the vibron bound state in CeCuGa$_3$ arises from the coupling of the first excited CEF level with the phonon mode, while in CeCuAl$_3$ it originates from the coupling of the second excited CEF level with the phonon mode.\cite{Adroja2012} This behavior can be understood as the first excited CEF level in CeCuGa$_3$ moves up in energy (compared to 1.3~meV in CeCuAl$_3$) and the phonon mode (due to heavy Ga atom compared to Al) moves down in energy. The CEF-phonon schematic for CeCuAl$_{3}$ \cite{Adroja2012} and CeAuAl$_{3}$ \cite{Petr2019} are also shown in Fig.~\ref{fig:CEF-phonon}. As seen from Fig.~\ref{fig:CEF-phonon}(c) the CEF-phonon energy level scheme for CeCuGa$_3$ is similar to that of CeAuAl$_{3}$.\cite{Petr2019}

Next, we calculate the crystal field susceptibility $\chi_{\rm CEF}(T)$ using the same CEF $B_n^m$ parameters as determined from the CEF-phonon model fit [Eq.~(\ref{H-CEF-ph})] of the INS data discussed above. Figure~\ref{fig:Chi_CEF} shows a comparison of the calculated $\chi_{\rm CEF}(T)$ with the experimental $\chi(T)$ for the polycrystalline and single crystal CeCuGa$_{3}$.  Figure~\ref{fig:Chi_CEF} reveals good agreement between the experimental and  calculated CEF-only (without phonon and MEL terms) susceptibility for both polycrystalline and single crystal of CeCuGa$_3$. The estimated value of the molecular field constant [in (mol/emu)] for the polycrystal sample $\lambda_{\rm p}=-22.1$, and for the single crystal $\lambda_{\rm a}=0.5$ and $\lambda_{\rm c}=103.4$. The values of the temperature independent susceptibility are [in ($10^{-4}$ emu/mole)], $\chi_{\rm 0p}= 1.2$,   $\chi_{\rm 0a}=-2.8$ and  $\chi_{\rm 0c}=-1.8$. The CEF-only simulation (without phonon and MEL terms) using the $B_{n}^{m}$ parameters obtained from the analysis of the INS data by the CEF-phonon model [Eq.~(\ref{H-CEF-ph})] suggests three CEF doublets at 0, 5.3 and 27.7~meV (see Fig.~\ref{fig:INS_sim_PCEEF} in Appendix). The wave functions of three CEF doublets are  $|\pm \frac{1}{2}\rangle $, $|\pm \frac{3}{2} \rangle$ and $|\pm \frac{5}{2} \rangle$. The values of the ground state moment estimated are $\mu_{x} = \mu_{y} =  1.28 \mu_B$ and $\mu_z$ =  0.43 $\mu_B$, which are in agreement with an easy plane ansiotropy observed through the single crystal susceptibility and neutron diffraction. The value of the ordered state moment determined from neutron diffraction at 1.7 K is 0.95(1)\,$\mu_{\rm B}$, a little smaller than the moment $\mu_{x} = \mu_{y} = 1.28\, \mu_{\rm B}$ obtained from the CEF ground state. This is likely due to the fact that the ordered moment is not yet saturated at 1.7 K as saturation is expected only at temperatures below $T_{\rm N}/2 \sim 1.2$~K\@.

Finally, it is worth noting that the presence of the anisotropic ferromagnetic correlations discussed in Sec.~\ref{MT-HC} is another important argument supporting the existence of a CEF-phonon coupling in CeCuGa$_{3}$. In fact, the CEF-phonon coupling is a manifestation of a highly anisotropic effective attraction between pairs of $4f$-type electrons ferromagnetically coupled via a phonon mode.\cite{Fuente2021} The presence of these highly anisotropic ferromagnetic correlations represents the clue that must be followed to find the CEF-phonon coupling in the antiferromagnetic systems like CeCuGa$_{3}$.

\section{\label{Concl} Conclusion}

We have performed a detailed examination of magnetism and crystal field effect in a Kondo lattice heavy fermion system CeCuGa$_{3}$ using the $\mu$SR, neutron powder diffraction and inelastic neutron scattering measurements. Both x-ray and neutron diffraction data consistently revealed a BaNiSn$_{3}$-type crystal structure (space group $I4\,mm$) for the polycrystalline CeCuGa$_{3}$ sample under study. The magnetic susceptibility and heat capacity reveal a magnetic phase transition near 2.3\,-\,2.5~K\@. The magnetic phase transition is further confirmed by the $ \mu$SR data which show a rapid loss in initial asymmetry in the magnetically ordered state. The nature of the magnetic phase transition was determined by neutron powder diffraction. The ND data reveal a long range antiferromagnetic ordering described by an incommensurate magnetic propagation vector {\bf k}~$ = (0.148, 0.148, 0)$. The magnetic structure is found to be a longitudinal spin density wave with a maximum ordered moment of $m = 0.95(1)\,\mu_{\rm B}$/Ce, and the ordered moments point along (1 1 0) direction. This magnetic structure is different from the one reported by Martin {\it et al}. \cite{Martin1998} who proposed a helical arrangement of moments. We notice that  along the $c$ axis the ordered moments are aligned ferromagnetically. As such, we attribute the indication of ferromagnetic coupling in heat capacity measured under applied field to this ferromagnetic alignment of ordered moments. From the magnetic heat capacity, magnetic entropy, Weiss temperature and quasielastic linewidth we estimate Kondo temperature to be 4\,-\,6~K\@.

\begin{figure}
\includegraphics[width=\columnwidth]{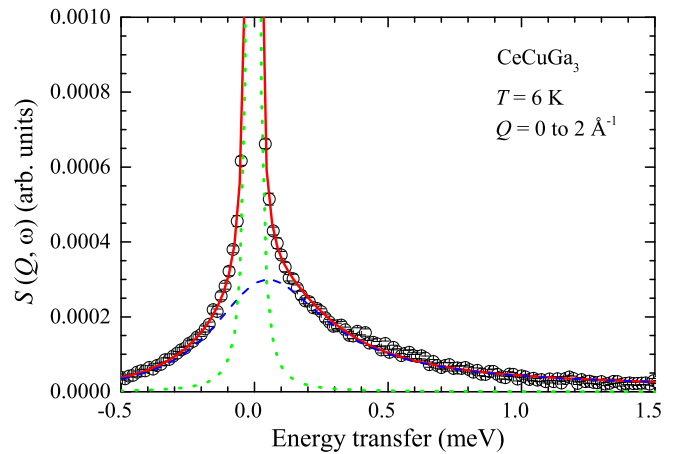}
\caption{\label{fig:INS-QENS} $Q$-integrated ($0 \leq Q \leq 2$~{\AA}$^{-1}$) low energy quasielastic scattering intensity $S(Q,\omega)$ versus energy transfer $E$ for CeCuGa$_{3}$ measured with final fixed energy of neutron, $E_{f}= 1.845$~meV at 6~K\@. The solid red line is the fit with a Lorentzian line-shape function for the quasielastic and elastic components. The dotted (elastic peak) and dashed (quasielastic peak) lines represent the components of the fit.}
\end{figure}

\begin{figure}
\includegraphics[width=6.5cm]{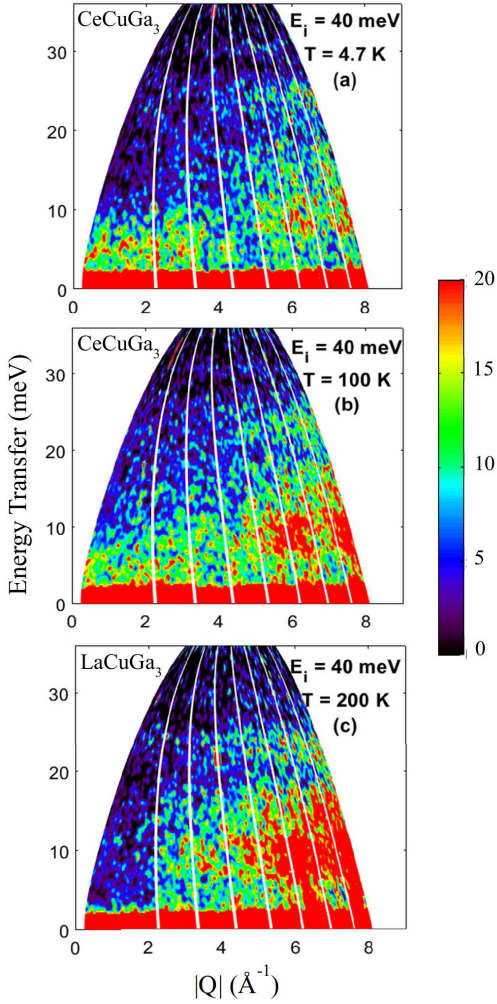}
\caption{\label{fig:contour-40meV} Inelastic neutron scattering response, a color-coded contour  map  of  the  intensity  (in  unit  of  mb/meV/sr/f.u.), energy  transfer $E$ versus  momentum  transfer $Q$ for (a) CeCuGa$_3$ at 4.7~K, (b) CeCuGa$_3$ at 100~K, and (c) LaCuGa$_3$ at 200~K (average temperature of the cooling run), measured with an incident  energy $E_i=  40$~meV.}
\end{figure}

Our investigation of the crystal electric field state using inelastic neutron scattering reveals two magnetic excitations near 4.5~meV and 6.9~meV and, possibility of a very broad and weak excitation between 20 and 30 meV. In the first model, the INS data were analyzed by a model based on pure crystal electric field, focusing on 4.5 and 6.9 meV excitations, and the crystal field level scheme was determined. However, our CEF level scheme deduced from the analysis of the INS data (4.5 and 6.9~meV excitations) is found to be substantially different from the one obtained from the analysis of the magnetic susceptibility of single crystal CeCuGa$_{3}$  ($I4/mmm$) by Joshi {\it et al}.\cite{Joshi2012} who found the two excited states at 4.3~meV and 19.6~meV. Our magnetic heat capacity data also support an overall splitting of 20.7~meV. Considering this, alternatively, in a second model, we analyzed the INS data (including 15 and 40 meV data) based on the CEF-phonon model which indicates that the two excitations at 4.5 meV and 6.9 meV have their origin in the CEF-phonon coupling (i.e. the splitting of one CEF peak into two peaks) observed in the case of homologue compounds CeCuAl$_{3}$ \cite{Adroja2012} and CeAuAl$_{3}$.\cite{Petr2019} The overall CEF splitting energy of 28.16~meV estimated from the INS analysis also explains the observed behavior of the magnetic heat capacity (magnetic entropy) as well as the single crystal magnetic susceptibility of CeCuGa$_3$. Further investigations, preferably on single crystal CeCuGa$_{3}$, are highly desired to examine the CEF-phonon coupling in this compound.

\vspace{0.5cm}
\noindent \textbf{Acknowledgments}
We thank Prof. S. Langridge, Prof. A Sundaresan, Dr. D. Le and Dr. Rotter for interesting discussions. DTA and VKA acknowledge financial assistance from CMPC-STFC grant number CMPC-09108. DTA would like to thank the Royal Society of London for Newton Advanced Fellowship funding between UK and China and International Exchange funding between UK and Japan. DTA also thanks EPSRC-UK for the funding. SS and RT would like to thank the Indian Nanomission (DST-India) for post-doctoral fellowship.

\section*{\label{Append} Appendix: Inelastic Neutron Scattering Data for low energy transfer  and ${\bm E_{i}= 40}$~\lowercase{me}V}

\begin{figure}
\includegraphics[width=\columnwidth]{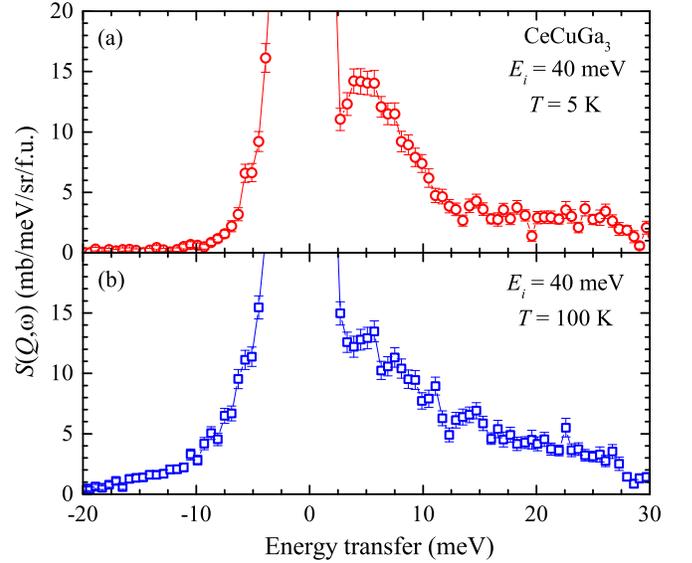}
\caption{\label{fig:INS-40meV} $Q$-integrated ($0\leq Q \leq 4$~{\AA}$^{-1}$) inelastic scattering intensity $S (Q,\omega)$ versus energy transfer $E$ for CeCuGa$_{3}$ measured with $E_{i}= 40$~meV at (a) 5~K, and (b) 100~K\@. }
\end{figure}

Figure~\ref{fig:INS-QENS} shows the $Q$-integrated ($0 \leq Q \leq 2$~{\AA}$^{-1}$) quasielastic neutron scattering data collected at 6~K using a PG002 analyser with final fixed energy of neutrons  $E_{f}= 1.845$~meV using the TOF spectrometer OSIRIS at the ISIS Facility. The quasielastic neutron scattering data were fitted by a Lorentzian line-shape function. From the fit we obtain the quasielastic line width $\Gamma_{\rm QE} = 0.517(2)$~meV, which yields a Kondo temperature $ T_{\rm K} = \Gamma_{\rm QE}/k_{\rm B}  = 6.0(3)$~K\@. The value of $ T_{\rm K}$ obtained this way is close to the values obtained from the heat capacity and Weiss temperature. 

\begin{figure}
\includegraphics[width=7cm]{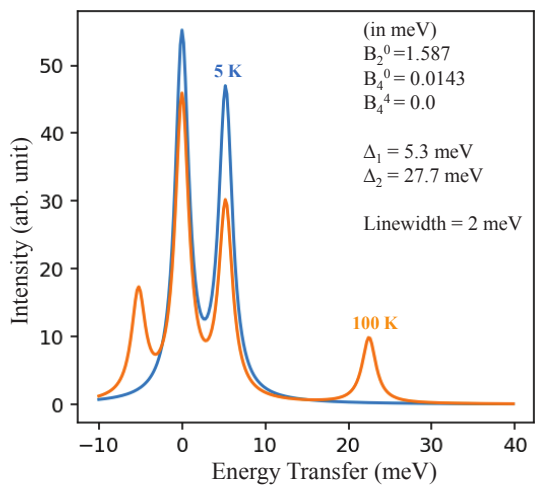}
\caption{\label{fig:INS_sim_PCEEF} Simulated inelastic scattering spectra at 5~K and 100~K using only the CEF $B_{n}^{m}$ parameters (without phonon and MEL terms) which are obtained from the analysis of the INS data by the CEF-phonon model [Eq.~(\ref{H-CEF-ph})].}
\end{figure}

The color-coded intensity maps for inelastic neutron scattering responses from CeCuGa$_{3}$ for $E_{i} = 40$~meV at $T = 4.7$~K and 100~K, and from LaCuGa$_{3}$ at 200~K are shown in Fig.~\ref{fig:contour-40meV}.  Figure~\ref{fig:INS-40meV} shows the $Q$-integrated ($0 \leq Q \leq 4$~{\AA}$^{-1}$) inelastic scattering intensity $S (Q,\omega)$ data which were collected with neutrons having $E_{i}= 40$~meV at 6~K using the TOF spectrometer MARI. As can be seen from Fig.~\ref{fig:INS-40meV} there is a very broad and weak excitation between  20 and 30~meV range of energy at 5~K\@.  This is in agreement with the heat capacity data, which suggest an overall CEF splitting of 20.7~meV.

Figure~\ref{fig:INS_sim_PCEEF} shows the CEF-only simulation (without phonon and MEL terms) using the CEF parameters $B_{2}^{0}$=1.587(5) meV, $B_{4}^{0}$=0.0143(2) meV and  $B_{4}^{4}$=0.000(1) meV, which are obtained from the analysis of the INS data by the CEF-phonon model [Eq.~(\ref{H-CEF-ph})]. As can be seen from the simulation, at 5~K there is an excitation from the ground state at 5.3~meV, giving $\Delta_1 = 5.3$~meV. At 5~K the CEF excitation from the ground state $|\pm 1/2 \rangle$ to the second excited state $|\pm 5/2\rangle$ (at 27.7~meV) is not allowed in the dipole approximation, and hence its intensity is zero in the INS simulation. However, the simulation for $T = 100$~K shows an excitation from the first excited state at 22.4~meV, which in turn implies $\Delta_2 = 22.4 + 5.3= 27.7$~meV (with respect to the ground state).


\begin{thebibliography}{98}


\bibitem{Stewart1984}
G. R. Stewart, Rev. Mod. Phys. {\bf 56}, 755 (1984); G. R. Stewart, Rev. Mod. Phys. {\bf 73}, 797 (2001); G. R. Stewart, Rev. Mod. Phys., {\bf 78}, (2006) 743.

\bibitem{Amato}
A. Amato, Rev. Mod. Phys. {\bf 69}, 1119 (1997).

\bibitem{Riseborough}
P. S. Riseborough, Adv. Phys. {\bf 49}, 257 (2000).

\bibitem{Lohneysen}
H. v. L\"{o}hneysen, A. Rosch, M. Vojta, and P. W\"{o}lfle, Rev. Mod. Phys. {\bf 79}, 1015 (2007).

\bibitem{Pfleiderer2009}
C. Pfleiderer, Rev. Mod. Phys. {\bf 81}, 1551 (2009).

\bibitem{Si2010}
Q. Si and F. Steglich, Science {\bf 329}, 1161 (2010);

\bibitem{Kimura2005}
N. Kimura, K. Ito, K. Saitoh, Y. Umeda, H. Aoki, and T. Terashima, Phys. Rev. Lett. {\bf 95}, 247004 (2005).

\bibitem{Kimura2007}
N. Kimura, Y. Muro, and H. Aoki, J. Phys. Soc. Jpn. {\bf 76}, 051010 (2007).

\bibitem{Sugitani2006}
I. Sugitani, Y. Okuda, H. Shishido, T. Yamada, A. Thamizhavel, E. Yamamoto, T. D. Matsuda, Y. Haga, T. Takeuchi, R. Settai, and Y. \=Onuki, J. Phys. Soc. Jpn. {\bf 75}, 043703 (2006).

\bibitem{Okuda2007}
Y. Okuda, Y. Miyauchi, Y. Ida, Y. Takeda, C. Tonohiro, Y. Oduchi, T. Yamada, N. D. Dung, T. D. Matsuda, Y. Haga, T. Takeuchi, M. Hagiwara, K. Kindo, H. Harima, K. Sugiyama, R. Settai, and Y. \=Onuki, J. Phys. Soc. Jpn. {\bf 76}, 044708 (2007).

\bibitem{Bauer2012}
E. Bauer and M. Sigrist (editors): Lecture Notes in Physics Vol. 847: \emph{Non-centrosymmetric Superconductors: Introduction and Overview}, (Spring-Verlag, Berlin Heidelberg, 2012).

\bibitem{Anand2011a}
V. K. Anand, A. D. Hillier, D. T. Adroja, A. M. Strydom, H. Michor, K. A. McEwen, and B. D. Rainford, Phys. Rev. B {\bf 83}, 064522 (2011).

\bibitem{Anand2011b}
V. K. Anand, D. T. Adroja, A. D. Hillier, J. Taylor, and G. And\'{r}e, Phys. Rev. B {\bf 84}, 064440 (2011).

\bibitem{Anand2011c}
 V. K. Anand, D. T. Adroja, A. D. Hillier, W. Kockelmann, A. Fraile, and A. M. Strydom, J. Phys.: Condens. Matter \textbf{23}, (2011) 276001.

\bibitem{Anand2012a}
V. K. Anand, D. T. Adroja, and A. D. Hillier, Phys. Rev. B {\bf 85}, 014418 (2012).

\bibitem{Adroja2012a}
D. T. Adroja and V. K. Anand, Phys. Rev. B {\bf 86}, 104404 (2012).

\bibitem{Hillier2012}
A. D. Hillier, D. T. Adroja, P. Manuel, V. K. Anand, J. W. Taylor, K. A. McEwen, B. D. Rainford and M. M. Koza, Phys. Rev. B {\bf 85},  134405 (2012). 

\bibitem{Anand2013}
V. K. Anand, D. T. Adroja, and A. D. Hillier, J. Phys.: Condens. Matter {\bf 25}, 196003 (2013).

\bibitem{Smidman2013}
M. Smidman, D. T. Adroja, A. D. Hillier, L. C. Chapon, J. W. Taylor, V. K. Anand, R. P. Singh, M. R. Lees, E. A. Goremychkin, M. M. Koza, V. V. Krishnamurthy, D. M. Paul and G. Balakrishnan, Phys. Rev. B {\bf 88}, 134416 (2013) . 

\bibitem{Anand2014a}
V. K. Anand, D. Britz, A. Bhattacharyya, D. T. Adroja, A. D. Hillier, A. M. Strydom, W. Kockelmann, B. D. Rainford and K. A. McEwen, Phys. Rev. B {\bf 90},  014513 (2014) . 

\bibitem{Anand2014b}
V. K. Anand, D. T. Adroja, A. Bhattacharyya, A. D. Hillier, J. W. Taylor and A. M. Strydom, J. Phys.: Condens. Matter {\bf 26}, 306001 (2014). 

\bibitem{Smidman2014}
M. Smidman, A. D. Hillier, D. T. Adroja, M. R. Lees, V. K. Anand, R. P. Singh, R. I. Smith, D. M. Paul, and G. Balakrishnan, Phys. Rev. B {\bf 89}, 094509 (2014).

\bibitem{Adroja2015}
D. T. Adroja, C. de la Fuente, A. Fraile, A. D. Hillier, A. Daoud-Aladine, W. Kockelmann, J. W. Taylor, M. M. Koza, E. Burzur\'{i}, F. Luis, J. I. Arnaudas, and A. del Moral Phys. Rev. B {\bf 91}, 134425 (2015). 

\bibitem{Anand2016}
V. K. Anand, D. T. Adroja, D. Britz, A. M. Strydom, J. W. Taylor and W. Kockelmann, Phys. Rev. B {\bf 94},  014440 (2016). 

\bibitem{Anand2018}
V. K. Anand, A. D. Hillier, D. T. Adroja, D. D. Khalyavin, P. Manuel, G. Andre, S. Rols, and M. M. Koza, Phys. Rev. B {\bf 97}, 184422 (2018). 

\bibitem{Adroja2012}
D. T. Adroja, A. del Moral, C. de la Fuente, A. Fraile, E. A. Goremychkin, J. W. Taylor, A. D. Hillier and F. Fernandez-Alonso, Phys. Rev. Lett. \textbf{108}, 216402 (2012).

\bibitem{Fuente2021} C. de la Fuente, A. del Moral, J.W. Taylor, D.T. Adroja, J. Magn. Magn. Mater. {\bf 530}, 167541 (2021).

\bibitem{Petr2019} P. \u{C}erm\'{a}k, A. Schneidewind, B. Liu, M. M. Koza, C. Franz, R. Sch\"{o}nmann, O. Sobolev, and C. Pfleiderer, Proc. Natl. Acad. Sci. {\bf 116}, 6695 (2019).

\bibitem{Sampathkumaran1992}
E. V. Sampathkumaran and I. Das, Solid State Commun. {\bf 81},  901 (1992).

\bibitem{Mentink1993}
 S. A. M. Mentink, N. M. Bos, B. J. van Rossum, G. J. Nieuwenhuys, J. A. Mydosh, and K. H. J. Buschow, J. Appl. Phys. \textbf{73}, 6625 (1993).

\bibitem{Martin1996}
Martin J M, Paul D M, Lees M R, Werner D and Bauer E, J. Magn. Magn. Mater. \textbf{159}, 223 (1996).

\bibitem{Aoyama1996}
S. Aoyama, H. Ido, T. Nishioka, M. Kontani, Czechoslov. J. Phys. {\bf 46}, 2069 (1996)

\bibitem{Martin1998}
J. M. Martin, M. R. Lees, D. M. Paul, P. Dai, C. Ritter, and Y. J. Bi, Phys. Rev. B {\bf 57}, 7419 (1998).

\bibitem{Kontani1999}
 Kontani M, Motoyama G, Nishioka T and Murase K: Physica B \textbf{259},  24 (1999).

\bibitem{Oe2010} K. Oe, Y. Kawamura, T. Nishioka, H. Kato, M. Matsumura and K. Kodama, J. Phys.: Conf. Ser. {\bf 200}, 012147 (2010).

\bibitem{Joshi2012}
 D. A. Joshi, P. Burger, P. Adelmann, D. Ernst, T. Wolf, K. Sparta, G. Roth, K. Grube, C. Meingast and H. v. L\"{o}hneysen,  Phys. Rev. B {\bf 86}, 035144 (2012). 
 
\bibitem{Przybylski2014}
P. Przybylski, A.P. Pikul, D. Kaczorowski, P. Wi\'sniewski, J. Phys. Chem. Solids {\bf 75}, 1284 (2014)



 \bibitem{Rodriguez1993}
J. Rodr\'{i}guez-Carvajal, Physica B {\bf 192}, 55 (1993); Program Fullprof, LLB-JRC, Laboratoire L\'{e}on Brillouin, CEA-Saclay, France, 1996 (\url{www.ill.eu/sites/fullprof/}).


 \bibitem{Gruner1974} G. Gr\"{u}ner and A. Zawadowski, Rep. Prog. Phys. {\bf 37}, 1497
(1974).

 \bibitem{Besnus1992}
M. J. Besnus, A. Braghta, N. Hamdaoui, and A. Meyer, J. Magn. Magn. Mater. {\bf 104-107}, 1385 (1992).

\bibitem{Blanco1994}
J. A. Blanco, M. de Podesta, J. I. Espeso, J. C. G\'{o}mez Sal, C. Lester, K. A. McEwen, N. Patrikios, and J. Rodr\'{i}guez Fern\'{a}ndez, Phys. Rev. B {\bf 49}, 15126 (1994).

\bibitem{Layek2009} S. Layek, V. K. Anand, and Z. Hossain, J. Magn. Magn. Mater.  {\bf 321}, 3447 (2009).

\bibitem{Prasad2012}	A. Prasad, V. K. Anand, U. B. Paramanik, Z. Hossain, R. Sarkar, N. Oeschler, M. Baenitz and C. Geibel,  Phys. Rev. B {\bf 86}, 014414 (2012). 

\bibitem{Rodriguez}
 J. Rodr\'{i}guez-Carvajal, BASIREPS: a program for calculating irreducible representations of space groups and basis functions for axial and polar vector properties. Part of the FullProf Suite of programs, \url{www.ill.eu/sites/fullprof/}

\bibitem{Ritter2011}
 C. Ritter, Solid State Phenomena 170, 263 (2011)

 \bibitem{Brown}
P. J. Brown, in International Tables for Crystallography, edited by A.J.C. Wilson, Mathematical, Physical and Chemical Tables, Vol. C (Kluwer Academic, Amsterdam, 1999), pp.450–457.

\bibitem{Stevens} K.\ W.\ H.\ Stevens,  Proc. Phys. Soc. London A {\bf 65}, 209 (1952).

\end{thebibliography}
\end{document}